\PassOptionsToPackage{unicode}{hyperref}
\PassOptionsToPackage{hyphens}{url}
\PassOptionsToPackage{dvipsnames,svgnames,x11names}{xcolor}
\documentclass[12pt]{article}

\usepackage{amsmath,amssymb}
\usepackage{iftex}
\ifPDFTeX
  \usepackage[T1]{fontenc}
  \usepackage[utf8]{inputenc}
  \usepackage{textcomp} 
\else 
  \usepackage{unicode-math}
  \defaultfontfeatures{Scale=MatchLowercase}
  \defaultfontfeatures[\rmfamily]{Ligatures=TeX,Scale=1}
\fi
\usepackage{lmodern}
\ifPDFTeX\else  
\fi
\IfFileExists{upquote.sty}{\usepackage{upquote}}{}
\IfFileExists{microtype.sty}{
  \usepackage[]{microtype}
  \UseMicrotypeSet[protrusion]{basicmath} 
}{}
\makeatletter
\@ifundefined{KOMAClassName}{
  \IfFileExists{parskip.sty}{%
    \usepackage{parskip}
  }{
    \setlength{\parindent}{0pt}
    \setlength{\parskip}{6pt plus 2pt minus 1pt}}
}{
  \KOMAoptions{parskip=half}}
\makeatother
\usepackage{xcolor}
\makeatletter
\ifx\paragraph\undefined\else
  \let\oldparagraph\paragraph
  \renewcommand{\paragraph}{
    \@ifstar
      \xxxParagraphStar
      \xxxParagraphNoStar
  }
  \newcommand{\xxxParagraphStar}[1]{\oldparagraph*{#1}\mbox{}}
  \newcommand{\xxxParagraphNoStar}[1]{\oldparagraph{#1}\mbox{}}
\fi
\ifx\subparagraph\undefined\else
  \let\oldsubparagraph\subparagraph
  \renewcommand{\subparagraph}{
    \@ifstar
      \xxxSubParagraphStar
      \xxxSubParagraphNoStar
  }
  \newcommand{\xxxSubParagraphStar}[1]{\oldsubparagraph*{#1}\mbox{}}
  \newcommand{\xxxSubParagraphNoStar}[1]{\oldsubparagraph{#1}\mbox{}}
\fi
\makeatother

\usepackage{longtable,booktabs,array}
\usepackage{calc} 
\usepackage{etoolbox}
\makeatletter
\patchcmd\longtable{\par}{\if@noskipsec\mbox{}\fi\par}{}{}
\makeatother
\IfFileExists{footnotehyper.sty}{\usepackage{footnotehyper}}{\usepackage{footnote}}
\makesavenoteenv{longtable}
\usepackage{graphicx}
\makeatletter
\def\maxwidth{\ifdim\Gin@nat@width>\linewidth\linewidth\else\Gin@nat@width\fi}
\def\maxheight{\ifdim\Gin@nat@height>\textheight\textheight\else\Gin@nat@height\fi}
\makeatother
\setkeys{Gin}{width=\maxwidth,height=\maxheight,keepaspectratio}
\makeatletter
\def\fps@figure{htbp}
\makeatother

\makeatletter
\@ifpackageloaded{caption}{}{\usepackage{caption}}
\AtBeginDocument{%
\ifdefined\contentsname
  \renewcommand*\contentsname{Table of contents}
\else
  \newcommand\contentsname{Table of contents}
\fi
\ifdefined\listfigurename
  \renewcommand*\listfigurename{List of Figures}
\else
  \newcommand\listfigurename{List of Figures}
\fi
\ifdefined\listtablename
  \renewcommand*\listtablename{List of Tables}
\else
  \newcommand\listtablename{List of Tables}
\fi
\ifdefined\figurename
  \renewcommand*\figurename{Figure}
\else
  \newcommand\figurename{Figure}
\fi
\ifdefined\tablename
  \renewcommand*\tablename{Table}
\else
  \newcommand\tablename{Table}
\fi
}
\@ifpackageloaded{float}{}{\usepackage{float}}
\floatstyle{ruled}
\@ifundefined{c@chapter}{\newfloat{codelisting}{h}{lop}}{\newfloat{codelisting}{h}{lop}[chapter]}
\floatname{codelisting}{Listing}

\makeatother
\makeatletter
\@ifpackageloaded{caption}{}{\usepackage{caption}}
\@ifpackageloaded{subcaption}{}{\usepackage{subcaption}}
\makeatother

\ifLuaTeX
  \usepackage{selnolig}  
\fi
\usepackage[]{natbib}
\bibpunct[, ]{(}{)}{,}{a}{,}{,}%
\usepackage{bookmark}

\IfFileExists{xurl.sty}{\usepackage{xurl}}{} 
\hypersetup{
  pdftitle={Title},
  pdfauthor={Author 1; Author 2},
  pdfkeywords={3 to 6 keywords, that do not appear in the title},
  colorlinks=true,
  linkcolor={blue},
  filecolor={Maroon},
  citecolor={Blue},
  urlcolor={Blue},
  pdfcreator={LaTeX via pandoc}}

\newcommand{\anon}{1}

\usepackage{NumericalInputs-v01}
\usepackage{NumericalFormats}
\usepackage{algorithm}
\usepackage{algpseudocode}
\usepackage{booktabs}
\usepackage{hyperref}
\usepackage{ifthen}
\usepackage{makecell}
\usepackage{pgf}
\usepackage{tikz}
\usepackage{xfp} 
\usepackage{enumitem}
\usepackage{changepage}

\usepackage{lineno}

\newcommand{\cbar}{\bar{c}}

\let\oldmarginpar\marginpar
\renewcommand\marginpar[1]{{\spacingset{1.0}\-\oldmarginpar[\raggedleft\scriptsize \textcolor{red}{#1}]%
	{\raggedright\scriptsize \textcolor{red}{#1}}}}

\newcommand{\significancemarker}[1]{%
  \ifthenelse{\lengthtest{#1 pt < 0.001 pt}}{\textsuperscript{***}}{%
    \ifthenelse{\lengthtest{#1 pt < 0.01 pt}}{\textsuperscript{**}}{%
      \ifthenelse{\lengthtest{#1 pt < 0.05 pt}}{\textsuperscript{*}}{%
        \ifthenelse{\lengthtest{#1 pt < 0.10 pt}}{\textsuperscript{$\dagger$}}{}}}}%
}

\newtheorem{assumption}{Assumption}
\newtheorem{remark}{Remark}

\newcommand{\sidebysidepicturewidht}{0.49}

\newcommand{\E}[1]{{\mathbb{E}}\left[#1\right]}

\newcommand{\ttheta}{\boldsymbol{\theta}}

\newcommand{\1}{\boldsymbol{1}}

\begin{document}

\def\spacingset#1{\renewcommand{\baselinestretch}%
{#1}\small\normalsize} \spacingset{1}


\newcommand{\papertitle}{Ballot Design and Electoral Outcomes: The Role of Candidate Order and Party Affiliation}

\if1\anon
{
  \title{\bf \papertitle}
  \author{Alessandro Arlotto\\
    The Fuqua School of Business, Duke University\\
    \href{mailto:aa249@duke.edu}{aa249@duke.edu}\bigskip\\    
    Alexandre Belloni\\
    The Fuqua School of Business, Duke University\\
    \href{mailto:abn5@duke.edu}{abn5@duke.edu}\bigskip\\    
    Fei Fang\\
    Yale School of Public Health, Yale University\\
    \href{mailto:fei.fang@yale.edu}{fei.fang@yale.edu}\bigskip\\    
    Sa\v{s}a Peke\v{c}\\
    The Fuqua School of Business, Duke University    \\    \href{mailto:pekec@duke.edu}{pekec@duke.edu}   }
  \date{}
  \maketitle
} \fi

\if0\anon
{
  \bigskip
  \bigskip
  \bigskip
  \begin{center}
    {\LARGE\bf \papertitle}
\end{center}
  \medskip
} \fi

\bigskip
\begin{abstract}
We develop a causal inference model to study how designing ballots with and without party designations impacts electoral outcomes when partisan voters rely on party-order cues to infer candidate affiliation in races without designations. 
If the party orders of candidates in races with and without party designations differ, these voters might cast their votes incorrectly.
We identify a quasi-randomized natural experiment with contest-level treatment assignment pertaining to North Carolina judicial elections and leverage double machine learning to accurately capture the magnitude of such incorrectly cast votes.
Using precinct-level election and demographic data, we estimate that {\DEMNonpartisanCubicVotingMistakesESTPercentage}
(SE: \DEMNonpartisanCubicVotingMistakesSEPercentage)
of Democratic partisan voters and
{\REPNonpartisanCubicVotingMistakesESTPercentage}
(SE: \REPNonpartisanCubicVotingMistakesSEPercentage)
of Republican partisan voters cast their votes incorrectly due to the difference in party orders.
A placebo test using judicial races with party designations shows that the flip effect disappears when party affiliation is observable on the ballot, which is consistent with the proposed causal mechanism.
\end{abstract}

\noindent%

{\it History:} First version: July 22, 2025; this version: July 21, 2026.
\vfill

\newpage
\spacingset{1.8} 

\section{Introduction}\label{se:introduction}

Elections rely on voters being able to correctly identify and vote for their preferred candidates. When voters make unintended choices, election outcomes may no longer accurately reflect underlying voter preferences.
Many factors shape how voters make decisions, including partisanship
\citep[see, e.g.,][]{Bartels:AmJofPoliticalSci2000,Fiorina:PB2002parties,BraderTucker:CompPolitics2012,Fiorina:Hoover2017}, incumbency
\citep[e.g.][]{GelmanKing:AJPS1990-Incumbency,KingGelman:AmJofPoliticalScience1991,CoxMorgenstern:LegislStudiesQuar1993,CoxKatz:AmJofPoliticalSci1996},
and campaign advertising
\citep[e.g.,][]{PhillipsUrbanyReynolds:JCR2008,WangLewisSchweidel:MarkSci2018-PoliticalAds,GordonLovettLuoReeder:MS2023=AdToneVoterTurnout}.

Yet voters' choices are shaped not only by political information and campaign dynamics but also by the design of the ballot itself.
Perhaps the best-known example is the butterfly ballot used in Palm Beach County, Florida, during the 2000 U.S. presidential election. Its unusual layout caused thousands of voters to select a different candidate than intended and may have affected the outcome of an election ultimately decided by a very narrow margin \citep[e.g.,][]{SinclairMarkMooreLavisSoldat:Nature2000electoral,WandShottsSekhon:AmPoliSciRev2001,Smith:StaSci2002}.
Although the butterfly ballot is an extreme case, a large literature shows that other aspects of ballot design can also influence voter behavior. These effects include ballot length and complexity, candidate order within contests, and the placement of contests on the ballot
\citep[e.g.,][]{MillerKrosnick:POQ1998,KoppellSteen:JPol2004,HoImai:JASA2006,HoImaiPOQ2008,KingLeigh:SSQ2009,MeredithSalant:PoliBe2013,AugenblickNicholson:REStud2015}. 

A wide variety of methods have been 
successfully applied to address a range of important election-related questions. 
These include, among others, election security 
\citep[e.g.,][]{FryOhlmann:Interfaces2009-VotingMachines,
HaseltineAlbert:WP2024-VotingByMail,
CrimminsHaldermanSturt:OPRE2025-VotingMachines}; 
the allocation and potential bias in the distribution of voting resources such as poll workers, polling places, and ballot drop boxes 
\citep[e.g.,][]{CachonKaaua:MS2022-VotingResources,
SchmidtBuellAlbert:MSOM2024-PollingLocations,
SchmidtAlbert:IISE2024-DropBox}; 
redistricting and gerrymandering
\citep[e.g.,][]{HessWeaverSiegfeldtWhelanZitlau:OPRE1965-redistricting,GelmanKing:APSR1994-Redistricting,
BernsteinDuchin:NofAMS2017-gerrymandering,
DeFordDuchinSolomon:HDSR2021-redistricting,
AutryCarterHerschlagHunterMattingly:MMS2021-Redistricting,
GurneeShmoys:ACDA2021-Redistricting,
BelottiBuchananEzazipour:OPRE2025-Redistricting,
ShahmizadBuchanan:OPRE2025-Redistricting, 
SwamyKingDouglasJacobson:OPRE2023-Redistricting,
ValidiBuchananLykhovyd:OPRE2022-Redistricting};
and polling methods and their potential biases
\citep[e.g.,][]{ShiraniMehrRothschildGoelGelman:JASA2018-BiasInPolls,GelmanHullmanWlezienElliottMorris:JDM2020,Gelman:SPP2021-PoliticalPolling,
GlassermanKuo:WP2025-VotingBias}.






In this paper, we focus on ballot design and
propose a causal inference framework to
study the effects of a ballot in which candidates' party designations (or labels) are included for certain \emph{partisan} contests and are omitted for other \emph{nonpartisan} contests.
Because voters' decisions are often strongly influenced by party identification in both such contests \citep{BonneauCann:PB2015},
the absence of party labels increases the chance that voters abstain from voting in those contests
\citep[also known as \emph{roll off}, cf.][]{EngstromCaridas:PPAD1991,VanderleeuwUtter:SSQ1993,AugenblickNicholson:REStud2015}
or resort to ballot cues 
\citep{Dubois:LawSocietyRev1984,EngstromCaridas:PPAD1991,McDermott:AmJPoliSci1997,MillerKrosnick:POQ1998,KoppellSteen:JPol2004,McDermott:JPol2005}.
In our analysis,  we focus on ballot cues and examine whether \emph{partisan voters}---those who vote by party even in nonpartisan contests---use the \emph{party order} from a contest with party designations (i.e., the sequence in which parties in that contest appear on the ballot) to inform their voting in nonpartisan contests.
If partisan voters cast their ballots assuming that the party orders of partisan and nonpartisan contests are the same, then their votes would not represent their intent when the two party orders actually differ.
We say that a \emph{(party-order) flip} occurs when the party order of a race differs from that of the U.S. presidential race in the same election. 
We are interested in measuring the impact of such a flip on nonpartisan vote shares,
and we postulate that such impact is \emph{heterogeneous} with respect to the presidential vote share. 
We are interested in two different estimands. 
The first estimand is the \emph{flip effect}, which measures the conditional average impact of a party-order flip on the vote share of a party’s candidate in a contest without party designations, 
given the vote share of the presidential candidate of the same party. 
The second estimand is the share of \emph{partisan-voting mistakes}, which measures the proportion of partisan voters whose vote is cast incorrectly due to the flip. 
In essence, partisan-voting mistakes represent the overall proportion of votes that do not reflect voter intent, while the flip effect measures the net impact on vote shares, considering that mistakes by opposing partisan voters cancel each other out.

We analyze {\nNonpartisanContests} North Carolina (NC) statewide judicial races \emph{without} party designations during the general elections of 2004, 2008, 2012, and 2016. 
Changes to the NC General Statute provide a quasi-randomized natural experiment with contest-level treatment assignment, 
allowing us to identify the causal effect of party-order flips.
By attributing party affiliations through contemporaneous endorsements, we find that {\nNonpartisanContestsFlipped} of these contests exhibit a party-order flip relative to the presidential race, while {\nNonpartisanContestsNotFlipped} do not. 
We construct a dataset for these {\nNonpartisanContests} contests containing a large number of electoral and demographic covariates and $\nNonpartisanObservationsOneParty$ observations. 
To accommodate the high dimensionality of the covariate space and allow for flexible functional forms, we employ double machine learning (DML), which combines machine learning predictive methods with valid statistical inference for the flip effect \citep[cf.][]{ChernozhukovChetverikovDemirerDufloHansenNeweyRobins:EconomJ2018,OprescuSyrgkanisBattocchiHEiLewis:NeurIPS2019econml}.
Furthermore, to account for the dependence of election outcomes within the same contest,
we compute cluster-robust standard errors.  
Finally, due to the heterogeneity of the effect across different levels of the presidential vote share, we use bootstrap to construct a simultaneously valid confidence band for the flip effect for all levels of the presidential vote share.

The remainder of the paper is organized as follows.
Section~\ref{se:Data&Motivation} describes the North Carolina statutory setting, the data, presents the results of preliminary analyses (including a simple OLS benchmark), and poses the scientific questions that motivate our analysis.
Section~\ref{se:Methodology} presents the causal inference framework, treatment model, and DML estimation procedure.
Section~\ref{se:results} reports the main results and placebo test.
Section~\ref{se:conclusions} concludes.

\subsection{Motivating example}\label{se:motivating-example}

\begin{figure}[t]
\linespread{1.25}\selectfont

\begin{adjustwidth}{-1cm}{-1cm}
\caption{\textbf{2016 North Carolina electoral maps by county}}
\label{fig:2016HeatMaps}
\centering{
 {\includegraphics[width=\sidebysidepicturewidht\linewidth]{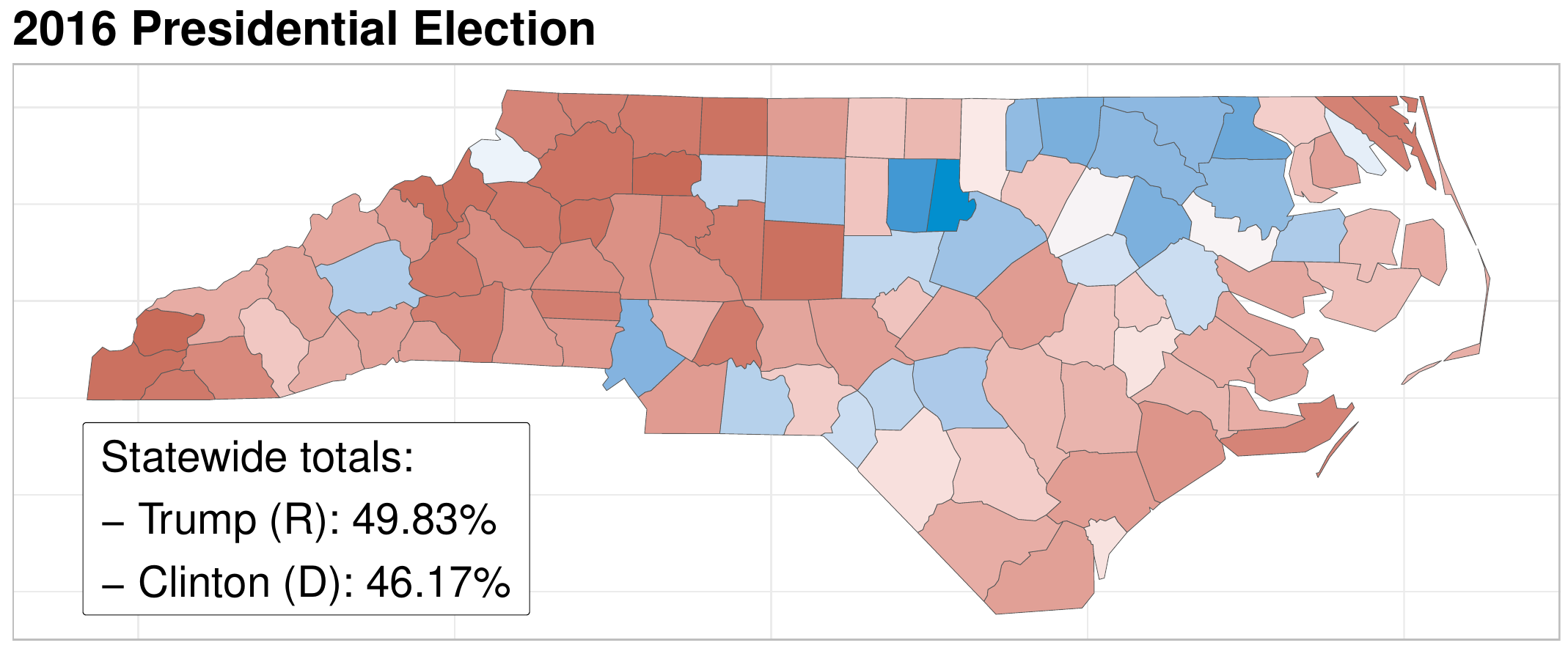}\hfill
  \includegraphics[width=\sidebysidepicturewidht\linewidth]{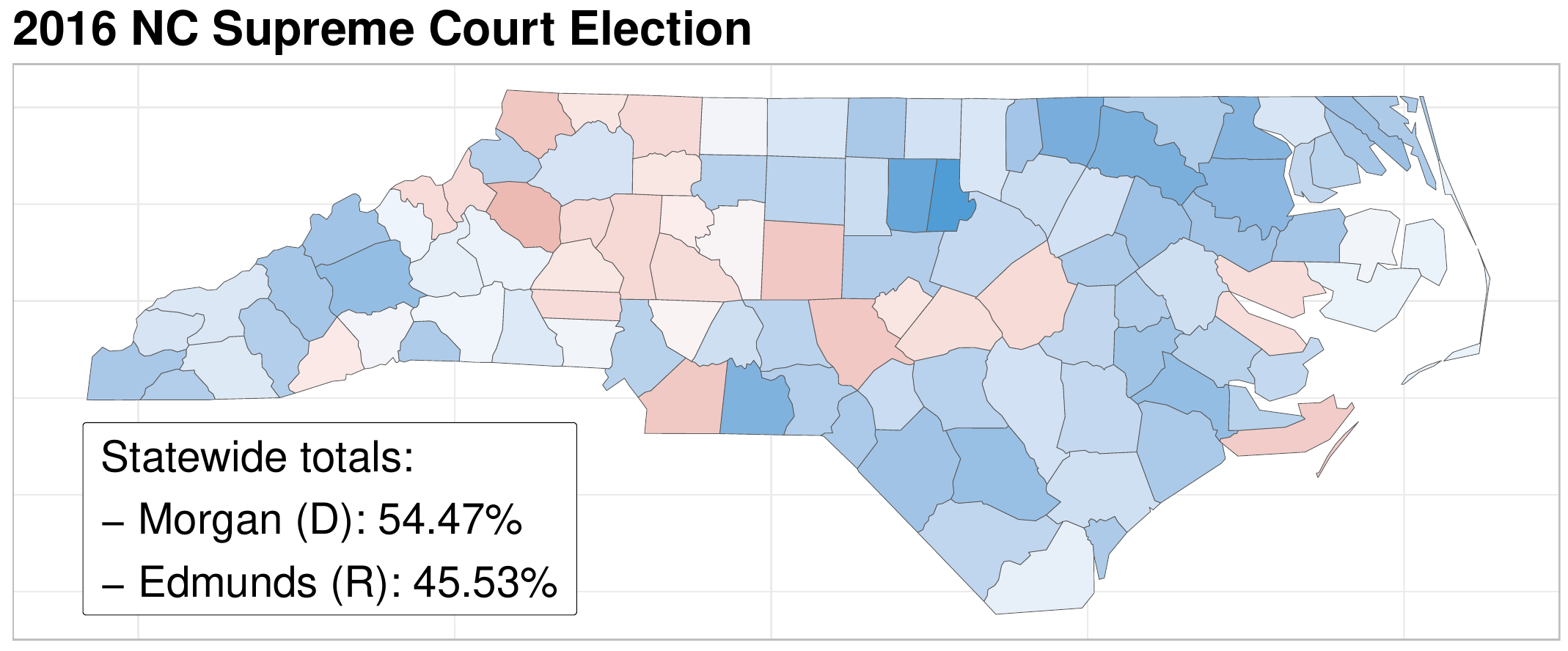}}}
 \caption*{\emph{Note.}
 The left panel displays the outcome of the 2016 presidential race in each of the 100 counties in North Carolina. 
 The Republican (Democratic) presidential candidate won the {\TrumpCountyWins} ({\ClintonCountyWins}) counties colored in red (blue), with darker shades indicating larger wins. 
 The right panel shows the analogous county outcomes of the 2016 NC Supreme Court race, using the same color coding.
 The 2016 NC Supreme Court race was on the same ballot as the 2016 presidential election.
 The presidential candidates were listed on the ballot \emph{with} their party labels: Donald J. Trump (R) was listed first, followed by Hillary Clinton (D). In contrast, the NC Supreme Court candidates were listed \emph{without} party labels. 
 Michael R. Morgan---endorsed by the Democrats---was listed first, while Robert H. Edmunds---endorsed by the Republicans---was listed second.
 Morgan won {\MorganCountyWins} counties in total, including {\MorganTrumpCountyWins} of the {\TrumpCountyWins} counties that Trump carried.
 }
\end{adjustwidth}
\end{figure}

The 2016 general election in North Carolina featured a key judicial contest for an Associate Justice seat on the NC Supreme Court. 
Heading into the election, the Republican justices on the court held a 4-3 majority, with the Republican incumbent Robert H. Edmunds running to remain in his seat and retain that majority. 
The challenger, Democratic judge Michael R. Morgan, was running for the same seat seeking to secure a Democratic majority on the court. 
In accordance with the NC General Statute (see Section \ref{se:arrangement-of-ballots}), the race was nonpartisan.
This meant that the names of the two candidates would appear on the ballot in a random order and without the corresponding party labels:
Judge Morgan (D) was listed first, and Justice Edmunds (R) was listed second.
At the time, the law also prescribed that candidates from the party of the sitting governor (Republican) would be listed first in all partisan statewide races, including the U.S. presidential race.
In the language of this paper, the \emph{party order} of the 2016 NC Supreme Court contest was \emph{flipped}, 
i.e., the party order of this nonpartisan judicial contest was the opposite of the party order of the presidential race. 

Figure \ref{fig:2016HeatMaps} presents the North Carolina county-level electoral maps of 
the 2016 U.S. presidential race (left panel)  
and of the 2016 NC Supreme Court race (right panel). 
The two maps show that, in most counties, the victory of a given party's presidential candidate corresponded to a loss or a weaker victory for the candidate of the same party running for the NC Supreme Court.
More specifically, the Republican presidential candidate Donald J. Trump carried {\TrumpCountyWins} of the 100 
counties in North Carolina (the counties shaded red in the left panel of Figure \ref{fig:2016HeatMaps}), while the Republican 
Supreme Court candidate Edmunds carried only {\EdmundsCountyWins} counties in total, 
corresponding to only {\EdmundsTrumpCountyWinsPct} of the counties 
carried by Trump. 
For comparison, the same election featured five Court of Appeals 
races with party labels. 
In each of these races, the Republican judicial candidates carried at least {\MinCOATrumpCountyWinsPct} 
of the counties won by Trump. 
This pattern is also reflected in precinct-level vote share correlations. 
The correlation between the precinct-level vote shares of Trump and Edmunds is {\TrumpEdmundsPrecinctCorrelation}, 
whereas the correlation between Trump's vote share and the vote share of each of the five Republican Court of Appeals candidates running with party labels is at least {\TrumpCOAMinPrecinctCorrelation}.
Analogous patterns hold for Democrats: the correlation between the precinct-level vote shares of Democratic presidential candidate Hillary Clinton and Morgan is {\ClintonMorganPrecinctCorrelation}, 
while the lowest correlation between Clinton's vote share and that of a Democratic Court of Appeals candidate running with a party label is {\ClintonCOAMinPrecinctCorrelation}.
The sharp drop in the correlation of the Supreme Court race is notable given 
that the Supreme Court race and the Court of Appeals races appeared on the same 
ballot and were voted on by the same electorate.
Taken together, these patterns are suggestive of a \emph{flip effect}: 
the party-order flip of the 2016 NC Supreme Court race 
may have influenced how some voters cast their ballots.

\section{Data and Motivation}\label{se:Data&Motivation}

In this section, 
we review the evolution of the NC General Statute regulating party designations and candidate orders (Section \ref{se:arrangement-of-ballots}), 
discuss the data used for our analysis (Section \ref{se:data}), 
present the results of preliminary analyses (Section \ref{se:preliminary-analysis}),
and pose our research question (Section \ref{se:research-question}).

\subsection{Study setting and quasi-randomized treatment variation}
\label{se:arrangement-of-ballots}

The NC General Statute, Chapter 163, \S165.5, 
determines whether a ballot should feature the party designations of the candidates in a specific contest, 
while \S165.6 dictates the order in which these candidates must be listed. 

For instance, during the general elections of 2004, 2008, and 2012, presidential candidates along with their party affiliations were arranged alphabetically by party (for the parties with a minimum 5\% statewide voter registration). 
Consequently, ballots featured the Democratic presidential candidate first, the Republican presidential second, followed by third-party candidates (in alphabetical order) and write-ins. 
In contrast, in those same elections, statewide judicial candidates were arranged alphabetically by last name, without party designation included on the ballot. 

Several amendments to the law were implemented leading up to the 2016 general election. 
Notably, party designation became mandatory for the candidates running for the NC Court of Appeals (Session Law, SL, 2015-292), and candidates of the party that received the most votes in the latest gubernatorial race would be listed first (SL 2013-381 and 2016-109).
Consequently, in 2016, the Republican presidential candidate appeared first on the ballot, and so did all of the Republican candidates for the NC Court of Appeals. 
In contrast, the candidates for Associate Justice of the NC Supreme Court lacked any party designation and were arranged alphabetically by last name, starting from the letter H.
This ballot arrangement was the result of two random drawings: 
the first drawing selected the letter H from a bowl, 
and the second drawing determined the order as alphabetical rather than reverse alphabetical.

Following the 2016 election, the NC General Assembly further amended the NC General Statute (SL 2016-125 and SL 2018-99) by 
(i) characterizing all statewide judicial elections as partisan, 
(ii) mandating their party designations to be printed on the ballot, 
(iii) and requiring all candidates in any election ballot to be listed in either alphabetical or reverse-alphabetical order, according to a drawing conducted by the State Board of Elections. 
In the 2020 general election, two random drawings determined candidates to be listed alphabetically, commencing with the  letter O; 
in 2024, the same process yielded 
alphabetical order starting from the letter D.

By regulating the order of candidates for each contest,
the NC General Statute establishes a corresponding party order,
either apparent with party labels or implied by party endorsements and affiliations.
The mechanism that determines the party order of a judicial contest has varied over time,
but the within-ballot order of candidates that such mechanisms induce is essentially randomized.
To see why, consider a given judicial contest with two candidates and take any two last names at random:
the arrangement rules give a 50\% chance
that either name appears first on that contest's ballot,
and because party labels are absent while each candidate is affiliated with one party,
this translates directly into a 50\% chance of a flip.
This quasi-randomized variation in party order provides the exogenous treatment variation
at the core of our identification strategy.

\subsection{Data}\label{se:data}

Our dataset consists of precinct-level data for {\nContests} statewide judicial races (NC Supreme Court and NC Court of Appeals) that took place during the general elections of 2004, 2008, 2012, 2016, 2020, and 2024.
Each of these races included two candidates from opposing parties
(a Republican and a Democrat), and we have a total of {\nObservationsOneParty} observations of judicial vote shares for each party.
During these elections, each ballot included between 15 and 20 statewide races, with the judicial races placed after all the other statewide races.
Our main causal analysis (Section~\ref{se:results-no-party-labels}) focuses on the {\nNonpartisanContests} statewide judicial races \emph{without} party designations from 2004, 2008, 2012, and 2016, and our placebo analysis (Section~\ref{se:results-yes-party-labels}) focuses on the {\nPartisanContests} statewide judicial races \emph{with} party designations from 2016, 2020, and 2024. The focus on presidential election years is driven by the fact that in North Carolina all major statewide offices are contested in such years, providing the richest set of covariates for our estimation procedure and the clearest identification of a party order from the top of the ticket.
Table~\ref{tab:summary-stats} provides a summary of the dataset, showing the number of judicial contests and precinct-level observations included in our analysis by year and treatment status, for each party.

\begin{table}[t!]
\linespread{1.25}\selectfont

\begin{adjustwidth}{-1cm}{-1cm}
\centering
\caption{\textbf{Statewide judicial races in North Carolina,
         presidential election years 2004--2024}}
\label{tab:summary-stats}
\footnotesize
\setlength{\tabcolsep}{12pt}
\begin{tabular}{l | r r r r | r r r r}
\toprule
  & \multicolumn{4}{c|}{\textbf{Without party labels (main sample)}}
  & \multicolumn{4}{c}{\textbf{With party labels (placebo sample)}} \\
\textbf{Year} 
  & \textit{Flip}
  & \textit{No flip}
  & \textit{Total}
  & \textit{Observations}
  & \textit{Flip}
  & \textit{No flip}
  & \textit{Total}
  & \textit{Observations} \\
\midrule
\textbf{2004} 
  & \FourNContestsFlipped
  & \FourNContestsNotFlipped
  & \FourNContests
  & \FourNObservations
  & & & & \\
\textbf{2008} 
  & \EightNContestsFlipped
  & \EightNContestsNotFlipped
  & \EightNContests
  & \EightNObservations
  & & & & \\
\textbf{2012} 
  & \TwelveNContestsFlipped
  & \TwelveNContestsNotFlipped
  & \TwelveNContests
  & \TwelveNObservations
  & & & & \\
\textbf{2016} 
  & \SixteenSCNContestsFlipped
  & \SixteenSCNContestsNotFlipped
  & \SixteenSCNContests
  & \SixteenSCNObservations
  & \SixteenCOANContestsFlipped
  & \SixteenCOANContestsNotFlipped
  & \SixteenCOANContests
  & \SixteenCOANObservations \\
\textbf{2020} 
  & & & &
  & \TwentyNContestsFlipped
  & \TwentyNContestsNotFlipped
  & \TwentyNContests
  & \TwentyNObservations \\
\textbf{2024} 
  & & & &
  & \TwentyFourNContestsFlipped
  & \TwentyFourNContestsNotFlipped
  & \TwentyFourNContests
  & \TwentyFourNObservations \\
\midrule
\textbf{Total}
  & \textbf{\nNonpartisanContestsFlipped}
  & \textbf{\nNonpartisanContestsNotFlipped}
  & \textbf{\nNonpartisanContests}
  & \textbf{\nNonpartisanObservationsOneParty}
  & \textbf{\nPartisanContestsFlipped}
  & \textbf{\nPartisanContestsNotFlipped}
  & \textbf{\nPartisanContests}
  & \textbf{\nPartisanObservationsOneParty} \\
\bottomrule
\end{tabular}
\caption*{\emph{Note.} The table reports the number of statewide judicial contests and precinct-level observations included in our analysis by year and treatment status (flipped vs.\ not flipped party order relative to the presidential race).}
\end{adjustwidth}
\end{table}

The dataset was built by combining precinct-level election results of statewide races with precinct-level information about NC voters who participated or were registered to vote in each election. This information is broken down by precinct, voting method, and voter demographics such as party affiliation, race, ethnicity, gender, and age.
The election and voter data we used is maintained by the NC State Board of Elections, and it is publicly available at \url{https://www.ncsbe.gov/}.
Additionally, we include election turnout statistics,
as well as the gender of each judicial candidate, 
whether they are listed first in their contest, if they are running as an incumbent, and their party affiliation. 

Because the law regulating the content and arrangement of official ballots has varied substantially over time (see Section \ref{se:arrangement-of-ballots}), 
our dataset includes {\nNonpartisanContests} judicial races without party designations and {\nPartisanContests} races with party designations. 
For the {\nNonpartisanContests} races \emph{without} party designations, 
we determine the party affiliation of each judicial candidate. 
We find {\nNonpartisanContestsFlipped} judicial races with a party-order flip relative to the presidential race and {\nNonpartisanContestsNotFlipped} without (totaling to $\nNonpartisanObservationsOneParty$ precinct-level vote shares per party). 
Among the {\nPartisanContests} races \emph{with} party designations, there are {\nPartisanContestsFlipped} that feature a party-order flip relative to the presidential race and {\nPartisanContestsNotFlipped} that do not (totaling to $\nPartisanObservationsOneParty$ vote shares per party). 
Thus, for each judicial race, we also include the binary treatment variable \emph{flip} to track whether the party order of the judicial race differs from the party order of the presidential race in the same election.

\subsection{Preliminary Analysis}\label{se:preliminary-analysis}

\begin{figure}[!t]
\linespread{1.25}\selectfont

\begin{adjustwidth}{-1cm}{-1cm}
 \caption{\textbf{Party-order flips and judicial vote shares in contests without party labels in North Carolina.}}
\centering{
 {\includegraphics[width=\sidebysidepicturewidht\linewidth]{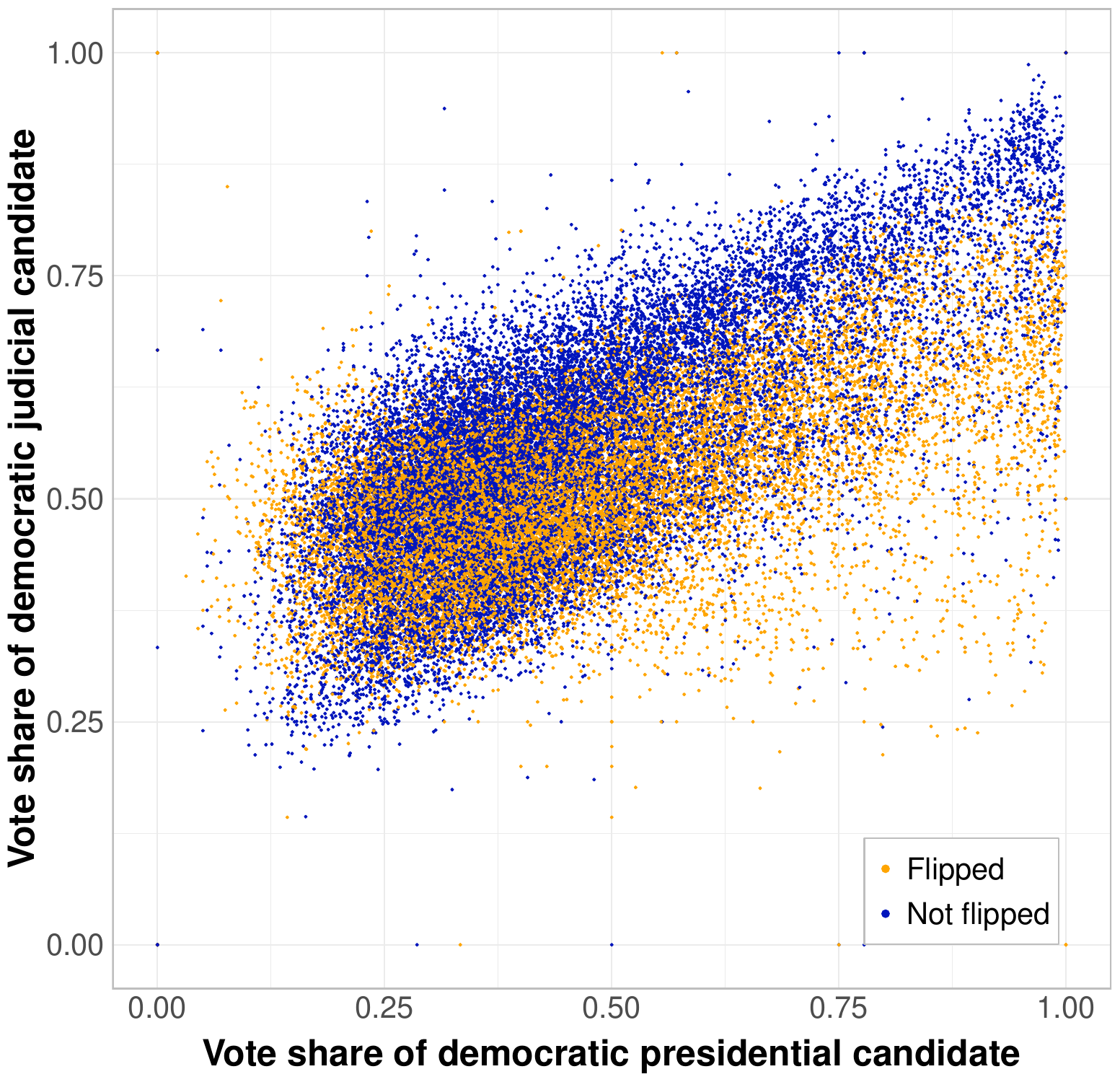}\hfill
 \includegraphics[width=\sidebysidepicturewidht\linewidth]
 {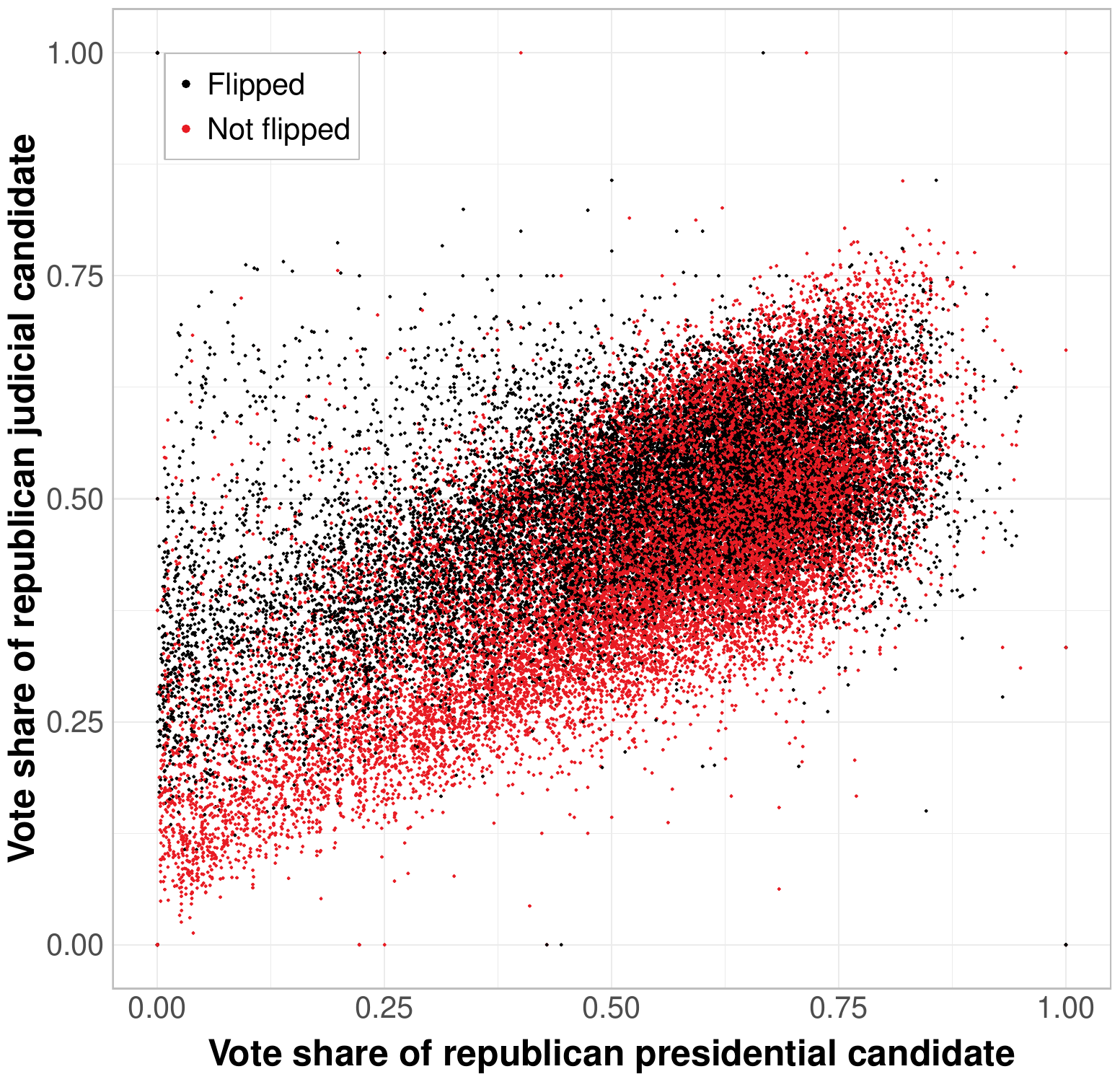}}}
 \caption*{\emph{Note.}
 The left panel shows {\nNonpartisanObservationsOneParty} precinct-level vote shares of Democratic judicial candidates ($y$-axis) who appeared on the ballot \emph{without} party designations during the 2004, 2008, 2012, and 2016 North Carolina general elections, plotted against the Democratic presidential vote share in the same precinct and election ($x$-axis).
 Judicial races with the same party order as the presidential race are shown in blue; those with flipped party order are shown in orange. 
 The  correlation between the Democratic judicial and presidential vote shares is $\DEMNonpartisanCorrelationPresidentialJudicialVoteSharesNONFlippedRaces$ in non-flipped races and $\DEMNonpartisanCorrelationPresidentialJudicialVoteSharesFlippedRaces$ in flipped races.
 The right panel presents the corresponding plot for Republican candidates during the same elections, with non-flipped judicial races in red and flipped judicial races in black.
 The  correlation between the Republican judicial and presidential vote shares is $\REPNonpartisanCorrelationPresidentialJudicialVoteSharesNONFlippedRaces$ in non-flipped races and $\REPNonpartisanCorrelationPresidentialJudicialVoteSharesFlippedRaces$ in flipped races.
 }
 \label{fig:FlippedNotFlippedScatterPlots} 
\end{adjustwidth}
\end{figure}

While the 2016 NC Supreme Court race of Section \ref{se:motivating-example} provided us with anecdotal evidence of a flip effect, 
a more comprehensive dataset with {\nNonpartisanObservationsOneParty} observations from judicial contests without party labels in the 2004, 2008, 2012, and 2016 general elections supports the same conclusion.
Figure \ref{fig:FlippedNotFlippedScatterPlots} 
plots the precinct-level vote share of a given party's judicial candidate against the precinct-level vote share of that party's  presidential candidate. By using different colors to distinguish judicial contests that are flipped from those that are not, we again see evidence of a flip effect.  
If the presidential candidate of a given party wins in a precinct, then the win of the judicial candidate of the same party tends to be weaker (and possibly a loss) if the party order of that judicial contest is flipped. 
Similarly, if the presidential candidate loses in a precinct, then the judicial candidate of the same party tends to lose by a smaller margin (or even wins) if the party order is flipped. 
As such, the effect of a party-order flip on the vote share of a judicial candidate would be positive when the vote share of the corresponding presidential candidate is low and negative when it is high.
In other words, the flip effect varies (is \emph{heterogeneous}) given the vote shares of the presidential contest. 

To corroborate this visual evidence, we compute the precinct-level  correlation between the judicial candidate's vote share and the presidential candidate's vote share of the same party, separately for flipped and non-flipped races. 
The correlations are substantially lower in flipped races than in non-flipped races for both parties. 
For Democratic candidates, the correlation drops from $\DEMNonpartisanCorrelationPresidentialJudicialVoteSharesNONFlippedRaces$ in non-flipped races to $\DEMNonpartisanCorrelationPresidentialJudicialVoteSharesFlippedRaces$ in flipped races; for Republican candidates, the corresponding figures are $\REPNonpartisanCorrelationPresidentialJudicialVoteSharesNONFlippedRaces$ and $\REPNonpartisanCorrelationPresidentialJudicialVoteSharesFlippedRaces$. 
This difference is consistent with the presence of a flip effect: 
when the party order of the judicial race is flipped relative to the presidential race, 
the link between how a precinct votes presidentially and how it votes judicially is weakened.

While the correlation analysis indicates that party-order flips affect voting behavior, it does not quantify the extent to which voters mistakenly support the judicial candidate of the opposite party.
As a starting point for quantifying the flip effect, we consider an ordinary least squares (OLS) interaction model.
If $c$ is a contest index and $p$ is a precinct index, then, for each party,
we let $Y_{cp}$ be the vote share of the party's judicial candidate in contest $c$ and precinct $p$,
$T_c$ be the treatment (flip) status of contest $c$, 
and $X_p$ be the vote share of the party's presidential candidate in precinct $p$.
Then, we consider the OLS model
\begin{equation}\label{eq:ols-interaction}
    Y_{cp} = \beta_0 + \beta_1 T_c + \beta_2 X_p + \beta_3 X_p T_c + \epsilon_{cp},
\end{equation}
separately for Democratic and Republican candidates with cluster-robust standard errors
at the contest level.
Evaluating model~\eqref{eq:ols-interaction} at $X_p = 1$ allows us to quantify partisan-voting mistakes by considering hypothetical precincts in which all voters support the same party in the presidential election, so that any voting mistake must come from those partisan voters.
The partisan-voting mistakes then are given by
$-(\beta_1+\beta_3)$ and capture the share of voters who vote against
their partisan preference when the ballot order is flipped.
For Democratic candidates, the estimated partisan-voting mistakes are
$\DEMNonpartisanOLSVotingMistakesESTPercentage$
(SE: $\DEMNonpartisanOLSVotingMistakesSEPercentage$);
for Republican candidates, the corresponding estimate is
$\REPNonpartisanOLSVotingMistakesESTPercentage$
(SE: $\REPNonpartisanOLSVotingMistakesSEPercentage$).

The OLS estimates of partisan-voting mistakes, however, lack precision.
The model also does not adjust for the many additional covariates available
(year fixed effects, contest type, ballot position, candidate gender, incumbency,
precinct-level vote shares from concurrent statewide races, and demographic and turnout data),
which are important both to reduce residual variance and to sharpen the estimates.
Adjusting for a large number of covariates requires regularization, 
which in turn introduces bias that invalidates standard OLS inference.
In Section~\ref{se:Methodology}, we describe how DML uses cross-fitting and the  orthogonality condition to debias the estimator of interest,
enabling valid inference even when machine learning methods are used
for covariate adjustment.

\subsection{Research Question}\label{se:research-question}

The scatter plots in Figure~\ref{fig:FlippedNotFlippedScatterPlots} suggest that flipping the party order weakens the link between judicial and presidential vote shares,
a pattern confirmed by the drop in  correlations between flipped and non-flipped races, as well as by the OLS interaction model \eqref{eq:ols-interaction}.
The same interaction model evaluated at a fully partisan precinct,
also provides a first estimate of the share of voters who cast their ballot
for the wrong candidate as a result of the flip.
These observations help us formulate the two research questions that guide the rest of the paper.
\begin{enumerate}[label=(\arabic*)]
    \item \emph{Is there a flip effect, and is it heterogeneous
    with respect to the presidential vote share?}
    The descriptive evidence above is suggestive, but a rigorous causal
    analysis is needed to establish this.
    \item \emph{What share of partisan voters cast their vote for the wrong candidate
    as a result of a party-order flip?}
    This is a key policy-relevant quantity that will influence how to design ballots.
\end{enumerate}

\section{Methodology} \label{se:Methodology}

In this section, we establish our causal inference framework (Section \ref{se:causal-inference-framework}) and discuss our estimation and inference procedures (Section \ref{se:estimation}).

\subsection{Causal inference framework, identification, and estimands}
\label{se:causal-inference-framework}

For each party (Democratic or Republican), we again consider the vote share $Y_{cp}$ of that party's judicial candidate in contest $c$ and precinct $p$.\footnote{The contest and precincts subscripts are both implicitly election year specific.}
We let $Y_{cp}(0)$ denote the potential vote share (i.e., potential outcome) when the party order of contest $c$ is the same as the party order of the presidential race on the same ballot, whereas $Y_{cp}(1)$ represents the potential vote share when the party order of contest $c$ is flipped. 
We use the binary variable $T_c \in \{0,1\}$ to denote the treatment status of contest $c$, which remains the same across all precincts in which contest $c$ appears, reflecting a contest-level treatment assignment. 
Specifically, $T_c = 1$ if the party order is flipped, 
and $T_c = 0$ otherwise.

In our analysis, we make the following standard assumptions for identification \citep[see, e.g.,][]{VanderWeele:StatMed2008}.
\begin{assumption}[No interference among units]
\label{asm:no-interference}
The precinct's potential vote shares of each given contest depend only on the treatment assigned to that contest and do not depend on the treatment assigned to other contests. 
\end{assumption}

\begin{assumption}[Known random assignment]
\label{asm:random-assignment}
The treatment is randomly assigned by a known mechanism. 
\end{assumption}

In the context of NC judicial elections without party labels, Assumption \ref{asm:no-interference} implies that the potential vote shares of a judicial candidate in one precinct do not depend on the party order of other judicial elections.
This assumption is likely to hold because voters either (i) have little or no visibility into the mechanisms that lead to different party orders on a ballot,
or (ii) already know which candidate they intend to support regardless of the party order of their contests. 
Such an assumption is common in many causal studies, and it has been successfully applied in related causal analyses of election ballot design
\citep[cf.][]{HoImai:JASA2006,AugenblickNicholson:REStud2015}.
Assumption~\ref{asm:random-assignment} is supported by the ballot arrangement rules of the NC General Statute, which induces quasi-random variation in party order across judicial contests and thereby provides exogenous treatment variation, as described in Section ~\ref{se:arrangement-of-ballots}.


We are interested in two (related) estimands. 
The first estimand is the \emph{flip effect}, defined as the conditional average treatment effect of a party-order flip on the vote share of a judicial candidate given the vote share of the presidential candidate from the same party and on the same ballot.
We expect the flip effect to be \emph{heterogeneous} given the presidential vote share.
When party designations are absent from the ballot, partisan voters may rely on the ordering of candidates as a cue to identify party affiliation, voting for the judicial candidate 
occupying the same position as their party's presidential candidate.
If the party order of the judicial contest is flipped relative to the presidential race, this 
behavior may lead some voters to cast a vote contrary to their intent.
Since outcomes are measured at the precinct level, 
the magnitude of this effect likely depends on the partisan composition of the precinct: 
in precincts with a larger share of one party's presidential voters, the incorrectly cast votes 
from that party's voters may be more numerous, shifting the judicial vote share against that 
party's candidate.
Conversely, in precincts with a smaller share of one party's presidential voters, the 
incorrectly cast votes from the opposing party's voters may dominate, shifting the judicial 
vote share in favor of the smaller-share party's candidate.
This suggests that the flip effect may change sign as the presidential vote share varies: it 
would be positive for low values of the presidential vote share of a given party and negative 
for high values. This is precisely the heterogeneity we estimate and test for.
Allowing for this type of heterogeneity is essential for accurately modeling precinct-level, down-ballot voting without party labels.
In our empirical analysis, we address these insights by formally testing (i) whether the flip effect is equal to zero and (ii) whether it is homogeneous.

The second estimand is the share of \emph{partisan-voting mistakes} attributable to such ballot design. 
This estimand aims to capture the intent of partisan voters, 
and is not directly associated with vote share outcomes. 
As a result, it could face an identification problem since 
we only observe the net effect on each outcome. 
However, 
the share of partisan-voting mistakes
can be identified by examining (hypothetical) precincts where the presidential candidate of a given party receives 100\% of the votes. 
In these precincts, there are no voters from the opposing party, eliminating the votes incorrectly cast by them.
Therefore, the identification of the (heterogeneous) flip effect (first estimand) suffices for the identification of the share of partisan-voting mistakes (second estimand) for such (hypothetical) precincts.

\subsection{Estimation and inference}\label{se:estimation}

To carry out our analysis, we consider $C$ judicial contests indexed by $c$, each with $n_c$ precincts, so that the total number of observations is $N = \sum_{c = 1}^{C} n_c$.
We let $X_p$ be the share of the vote in the precinct $p$ for the presidential candidate on the same ballot and of the same party as the candidate in contest $c$.
For each contest $c \in   [C] \equiv \{1, 2, \ldots, C\}$ and precinct $p \in 
\{1, 2, \ldots, n_c\}$,
we also let $W_{cp}$ and $Z_c$, respectively, be the precinct- and contest-level covariates.

To implement DML, we specify an \emph{outcome model} that allows the flip effect to vary with the presidential vote share, and a \emph{treatment model} that exploits the quasi-randomized assignment of party-order flips at the contest level.
For the outcome model, we set:
\begin{equation}\label{eq:outcome-model}
    Y_{cp} =  f(X_p) T_c + g(X_p, W_{cp}, Z_c) + \epsilon_{cp},     
\end{equation}
where $f$ denotes the \emph{flip effect}, the conditional  average treatment effect of a party-order flip given
the presidential vote share $X_p$.
In \eqref{eq:outcome-model}, the nuisance function $g$ 
captures the contribution of $X_p$ and the other covariates when the party order is not flipped,
and the error terms $\epsilon_{cp}$ are assumed to be independent across contests and centered, i.e., $\E{\epsilon_{cp} | T_c, X_p, W_{cp}, Z_c} = 0$. 

We consider a specification in which the flip effect $f$ is modeled as a non-linear function of the presidential vote share, specifically as a polynomial of degree $q \geq 0$: 
\begin{equation}\label{eq:f-qpoly-specification}
f(x) 
= \theta_0  + \theta_1  x + \theta_2  x^2 + \cdots + \theta_q x^q,
\end{equation} 
where $x$ is the presidential vote share and $\ttheta = (\theta_0, \ldots, \theta_q)$ is the vector of  coefficients to be estimated.

For the treatment model, we set:
\begin{equation}\label{eq:treatment-model}
    T_c = m_0 + U_c, 
\end{equation}
where $m_0 = \mathbb{E}[T_c]$ is the probability of being treated (i.e., the party order of the judicial contest is flipped relative to that of the presidential race) 
and $U_c$ is a mean-zero contest-level error term that is independent of $(X_p, W_{cp}, Z_c)$.
The choice of specification \eqref{eq:treatment-model} is motivated by our empirical setting. 
First, the NC General Statute quasi-randomizes the treatment (party-order flip) at the contest level, as detailed in Section~\ref{se:arrangement-of-ballots}, which implies that 
$\mathbb{E}[T_c \mid X_p, W_{cp}, Z_c] = \mathbb{E}[T_c]$ is constant, 
so the treatment is independent of all covariates.
As a result, the general form of the 
treatment model, $T_c = m_0(X_p, W_{cp}, Z_c) + U_c$, where $m_0$ is a potentially flexible function of the covariates, reduces to \eqref{eq:treatment-model}.
Second, the treatment is assigned at the contest level, so the effective sample size for 
estimating a flexible $m_0$ is the number of contests $C$, which is small, reducing any 
potential efficiency gain from modeling $m_0$ flexibly.

In our analysis, we estimate the flip effect across the full range of presidential vote shares and formally test several hypotheses regarding its structure. 
The first test, which we refer to as the \emph{test of zero flip effect}, evaluates whether party-order flips have any effect on judicial election outcomes.
Conceptually, this test corresponds to the null hypothesis 
$H^z_0: f(x) = 0  \text{ for all } x \in [0,1]$.\footnote{Appendix \ref{appendix:AlternativeTests} discusses alternative tests that can be implemented via bootstrap.}
Under the polynomial specification in \eqref{eq:f-qpoly-specification}, 
we implement this by testing the joint hypothesis 
\begin{equation}\label{test:ZeroFlipEffect}
H^z_0: \theta_0=\theta_1=\cdots=\theta_q = 0.
\end{equation}

The second test, which we refer to as the \emph{test of homogeneous flip effect}, 
examines whether the flip effect remains constant across all values  of the presidential vote share,
$x \in [0,1]$.
The corresponding null hypothesis is
$H^h_0$: there exists $\cbar$ such that $f(x) = \cbar$  for all  $x \in [0,1]$.
As with the previous test, under the polynomial specification in \eqref{eq:f-qpoly-specification},
this hypothesis can be assessed by testing restrictions on polynomial coefficients. 
Specifically, in our analysis, we test the joint hypothesis
\begin{equation}\label{test:HomogeneousFlipEffect}
H^h_0: \theta_1=\theta_2=\cdots=\theta_q = 0.
\end{equation}

We next describe how we use DML to estimate the flip effect and test the corresponding hypothesis.
Based on \eqref{eq:treatment-model}, we estimate $m_0$ by the sample mean 
$\hat m_0 = \frac{1}{C}\sum_{c=1}^C T_c$.
With the specifications in \eqref{eq:outcome-model} and \eqref{eq:f-qpoly-specification},
we then apply double machine learning
(see \citealp[Section 6.1]{ChernozhukovChetverikovDemirerDufloHansenNeweyRobins:EconomJ2018}, 
and \citealp{chernozhukov2018generic}) 
to estimate the flip effect for a given polynomial degree $q$.
Specifically, we estimate the vector of polynomial coefficients $\hat \ttheta = (\hat \theta_0, \ldots, \hat \theta_q)$, and use it to construct the point estimates 
$\hat f(x) = \hat\theta_0 + \hat\theta_1 x + \cdots + \hat\theta_q x^q$ for all $x \in [0,1]$. 
The corresponding point estimate for the fraction of partisan-voting mistakes, $\tau$, is then given by 
\begin{equation}\label{eq:partisan-voting-mistakes}
\hat \tau = - \hat f(1)  = - ( \hat\theta_0 + \hat\theta_1+\cdots+ \hat\theta_q ).
\end{equation}
To estimate the nuisance function $g$ in the outcome model \eqref{eq:outcome-model}, we use Elastic Net with cross-validated penalty selection \citep{HuiHastie:JRSSB2005,FriedmanHastieTibshirani:JSS2010}, 
implemented through the \texttt{ElasticNetCV} estimator provided by the \texttt{EconML} Python package \citep{EconML}. 
This approach handles a large number of covariates while accommodating a flexible functional form.
The covariates are detailed in Appendix~\ref{appendix:variables} and include characteristics of 
each judicial race (year fixed effects, contest type, contest position on the ballot, 
candidate position in contest, candidate gender, incumbency and party status), 
features pertaining to concurrent statewide races (precinct-level vote shares and winning 
indicators from presidential, gubernatorial, lieutenant governor, auditor, commissioner of 
agriculture, commissioner of insurance, commissioner of labor, secretary of state, 
superintendent of public instruction, and treasurer races), as well as precinct-level cross-tabulated
demographic data (age, party registration, race, ethnicity, and gender) of registered and 
actual voters, voting methods, and turnout.
In total, these amount to more than {\NCovariates} covariates.
By adjusting for this rich set of covariates, the DML estimator reduces residual variation and thereby increases the precision of the estimated flip effect coefficients $\hat\ttheta$.
Crucially, the orthogonality condition underlying DML preserves root-$N$ consistency of our estimator $\hat{\ttheta}$ and the validity of our confidence regions and p-values.
This efficiency gain is valuable even under quasi-randomized assignment, 
where covariate adjustment in the treatment model is not required for consistency 
of the flip effect estimator, 
but has the potential to tighten the confidence intervals 
for the conditional average treatment effect at each value of the presidential 
vote share.

Algorithm~\ref{alg:dml-estimation} outlines the DML 
estimation steps in detail, with particular emphasis on our setting, which involves both randomized and clustered treatment assignment.
Algorithm~\ref{alg:dml-estimation} computes residuals $\hat V$ from the machine learning outcome model 
\eqref{eq:outcome-model} and (contest-constant) residuals $\hat U$ from the treatment model 
\eqref{eq:treatment-model}, then regresses $\hat V$ on $\hat U$ to estimate 
the flip effect coefficients $\hat{\ttheta} = (\hat\theta_0, \hat\theta_1, 
\ldots, \hat\theta_q)$ in \eqref{eq:f-qpoly-specification} and to compute the corresponding residuals $\hat \varepsilon$,
with $\hat \varepsilon_c $ denoting the vector of residuals for cluster $c$.
At any presidential vote share $x \in [0,1]$, the estimated flip effect is then 
given by the polynomial
$\hat{f}(x) = \hat\theta_0 + \hat\theta_1 x + \hat\theta_2 x^2 + \cdots + 
\hat\theta_q x^q.$

\begin{algorithm}[t!]
\footnotesize
\caption{DML with randomized and clustered treatment assignment}
\label{alg:dml-estimation}
\begin{algorithmic}[1]

\State \textbf{Input:} Observations $(Y_{cp}, T_c, X_p, W_{cp}, Z_c)$ for contests (clusters) $c \in [C]$ and precincts $p \in [n_c]$.

\State \textbf{Outcome-model residuals:}
Estimate the function $g$ by $\hat g$ using \texttt{ElasticNetCV} with cross-fitting
and compute the residuals
$$
\hat V_{cp} = Y_{cp} - \hat{g}(X_p, W_{cp}, Z_c).$$
Let $\hat V_c = (\hat V_{c1}, \ldots, \hat V_{cn_c})^T$
denote the vector of residuals for cluster $c$,
and set $\hat V = (\hat V_1^T, \ldots, \hat V_C^T)^T$.

\State \textbf{Treatment-model residuals:}
Estimate the randomized and clustered treatment using its sample mean
and compute the residual
\[
\hat U_{c}
=
T_c-\frac{1}{C}\sum_{c=1}^C T_c.
\]
Let $\mathbf{1}_{n_c}$ be an $n_c$-dimensional vector of ones, and 
define $\hat U = (\hat U_1 \1_{n_1}^T,\ldots,\hat U_C \1_{n_C}^T)^T$.

\State \textbf{Estimation of treatment effect:}
Regress $\hat V_{cp}$ on  $ (X^0_{p} \hat U_{c}, X^1_{p} \hat U_{c},\ldots, X^q_{p} \hat U_{c})$ 
to estimate
$$
\hat \ttheta = (\hat \theta_0, \ldots, \hat \theta_q)
\qquad \text{and} \qquad
\hat \varepsilon_{cp} = \hat V_{cp} - \sum_{i=0}^{q} \hat{\theta}_i X^{i}_{p} \hat U_{c}.
$$
Let $\hat \varepsilon_c = (\hat \varepsilon_{c1}, \ldots, \hat \varepsilon_{cn_c})^T$ denote the vector of residuals for cluster $c$
and set $\hat \varepsilon = (\hat \varepsilon_{1}^T, \ldots, \hat\varepsilon_C^T)^T$.

\State \textbf{Output:}
Coefficient estimates $\hat{\ttheta}$, and residuals $\hat V,  \hat U, \hat \varepsilon$.

\end{algorithmic}
\end{algorithm}

\begin{algorithm}[t!]
\footnotesize
\caption{Adjusted multiplier bootstrap for uniform critical values}
\label{alg:Bootstrap}
\begin{algorithmic}[1]

\State \textbf{Input:}
Sample size $N$,
scaling matrix $J$,
scores $\Psi^i_c$ for $i \in \{0, 1, \ldots, q\}$ and $c \in [C]$, cluster-robust variance $\hat{\sigma}^2(\hat{f}(x))$,
replications $M$,
significance level $\alpha$.

\For{$ m \in  [M] $}

        \State  \textbf{Weights:} Sample $\zeta_{cm} \sim \mathcal{N}(0, 1) $ independently for $c \in [C]$. Set $\zeta_m =(\zeta_{1m},\ldots, \zeta_{Cm})^T$.

        \State \parbox[t]{\dimexpr\linewidth-\algorithmicindent}{\textbf{Multiplier statistic:}
        Given $\Psi^i_c$  for $i \in \{0, 1, \ldots, q\}$ and $c \in [C]$,
        define $\Psi^i = (\sum_{j=1}^{n_1}\Psi^i_{1j}, \cdots, \sum_{j=1}^{n_C}\Psi^i_{Cj})^T$, and
        construct the matrix $\Psi = (\Psi^0, \ldots, \Psi^{q})$. \\
        For each $x \in [0,1]$, compute the statistic
        $$
        t_{m}(x) = \frac{1}{N \hat{\sigma}(\hat{f}(x))} (1, x, x^2, \ldots, x^q) J^{-1} {\Psi}^T \zeta_m.
        $$}
\EndFor

    \State Conditional on the data compute the critical value
    $$
    q^*_{1-\alpha} = (1-\alpha)\mbox{-quantile  of } \left\{ \sup_{x \in [0,1]} |t_m(x)| : m \in [M] \right\}.
    $$

\State\textbf{Output:} Uniform critical value, $q^*_{1-\alpha}$.
\end{algorithmic}
\end{algorithm}

Moreover, because precinct-level outcomes are correlated within the same judicial contest, 
we account for this intra-cluster dependence by computing cluster-robust standard errors for both
$\hat \ttheta$ and $\hat f(x)$, $x \in [0,1]$. 
For each $i \in \{0,1, \ldots, q\}$ and $c \in [C]$, 
we define $X_c^i = (X^i_{p})_{p=1}^{n_c}$ 
to be the (column) vector of $i$th powers of the presidential vote shares pertaining to contest $c$, and we set $X^i = ({X_1^i}^T, \ldots, {X_C^i}^T)^T$.
Let also $\mathbf{1}_{n_c}$ be an $n_c$-dimensional vector of ones.
Using these vectors, 
we compute the 
scores
$
\Psi^i_c = X^i_c \circ {\hat U_c \1_{n_c}} \circ {\hat \varepsilon}_c,
$ 
where $\circ$ denotes the elementwise (Hadamard) product.
Then, we define the 
scaling matrix  $J$ as
$$
J = \frac{1}{N}
({X}^0 \circ {\hat U}, {X}^1 \circ {\hat U}, \ldots, {X}^q \circ {\hat U})^T
({X}^0 \circ {\hat U}, {X}^1 \circ {\hat U}, \ldots, {X}^q \circ {\hat U}).
$$
For each contest $c\in[C]$, let $A_c$ denote the matrix whose $(i,j)$th entry is 
$
(A_c)_{ij}
=
\sum_{k=1}^{n_c}\sum_{\ell=1}^{n_c}
(\Psi_c^{\,i}\Psi_c^{\,jT})_{k\ell},
$ 
for $i,j=0,\ldots,q$.
Then, the cluster-robust variance for $\hat{\ttheta}$ is given by
\citep[see, e.g.,][]{cameron2015practitioner}
\[
\widehat{\mathrm{Var}}(\hat{\ttheta})
=
\frac{1}{N^2}
J^{-1}
\left(
\sum_{c=1}^C A_c
\right)
J^{-1}.
\]
Given this, the cluster-robust variance of the estimated flip effect at any presidential vote share $x \in [0,1]$ is
$$
\hat{\sigma}^2(\hat{f}(x))
= (1, x, x^2, \ldots, x^q) \widehat{\text{Var}}(\hat{\ttheta}) (1, x, x^2, \ldots, x^q)^T.
$$
This variance estimator is used to construct the pointwise confidence bands shown in Figures~\ref{fig:AverageTreatmentEffectNoPartyLabels} and~\ref{fig:AverageTreatmentEffectWPartyLabels}.
It is also used in the bootstrap procedure described in Algorithm~\ref{alg:Bootstrap} to compute the critical values for the uniform confidence bands displayed in the same figures \citep[see also][]{ChiangKatoMaSasaki:JBusEconStat2022},
and for the hypothesis tests reported in this paper.

\begin{remark}[Party-specific estimation]\label{rem:parti-specific-estimation}
In implementing the estimation procedure described in this section, we estimate the flip effect separately for Democratic and Republican candidates. 
In two-candidate judicial races, Democratic and Republican vote shares sum to one by construction, inducing perfect negative dependence between the two outcome variables. 
Joint estimation would therefore require explicitly accounting for this dependence structure in the DML score and in the construction of cluster-robust standard errors. 
Estimating separately by party avoids these complications. 
It is also consistent with empirical work in political science that estimates party- or candidate-specific effects using separate models  \citep[see, e.g.,][]{HoImaiPOQ2008,BailArgyleBrownBumpusChenHunzakerLeeMannMerhoutVolfovsky-PNAS2018-PoliticalPolarization,Kousser-Phillips-Shor-PSRM-2018-Reform&Representation,MacInnisMillerKrosnickBelowLindner:PLoS2021}.
Furthermore, we do not impose any symmetry or other cross-party constraints linking the Democratic and Republican flip effects.
\end{remark}

\section{Results from North Carolina's Judicial Elections}\label{se:results}

We now use our dataset of North Carolina judicial elections to estimate the effect of a flip in the party order of judicial candidates relative to the party order of presidential candidates.
Section \ref{se:results-no-party-labels} presents our main results, based on data from judicial races \emph{without} party designations and on a cubic specification of the flip effect. The corresponding results obtained under the linear and constant specifications of the flip effect are reported in Appendix~\ref{se:linear&constant-specifications}.
Section \ref{se:results-yes-party-labels} uses data from judicial races \emph{with} party designations to conduct a placebo test.

\subsection{Main result: conditional average flip effect without party labels (cubic specification)}\label{se:results-no-party-labels}

{

\begin{figure}[p]
\linespread{1.25}\selectfont

\caption{\textbf{Flip effect without party labels (cubic specification)}}
\label{fig:AverageTreatmentEffectNoPartyLabels}

\begin{adjustwidth}{-1cm}{-1cm}
\centering

{\includegraphics[width=\sidebysidepicturewidht\linewidth]{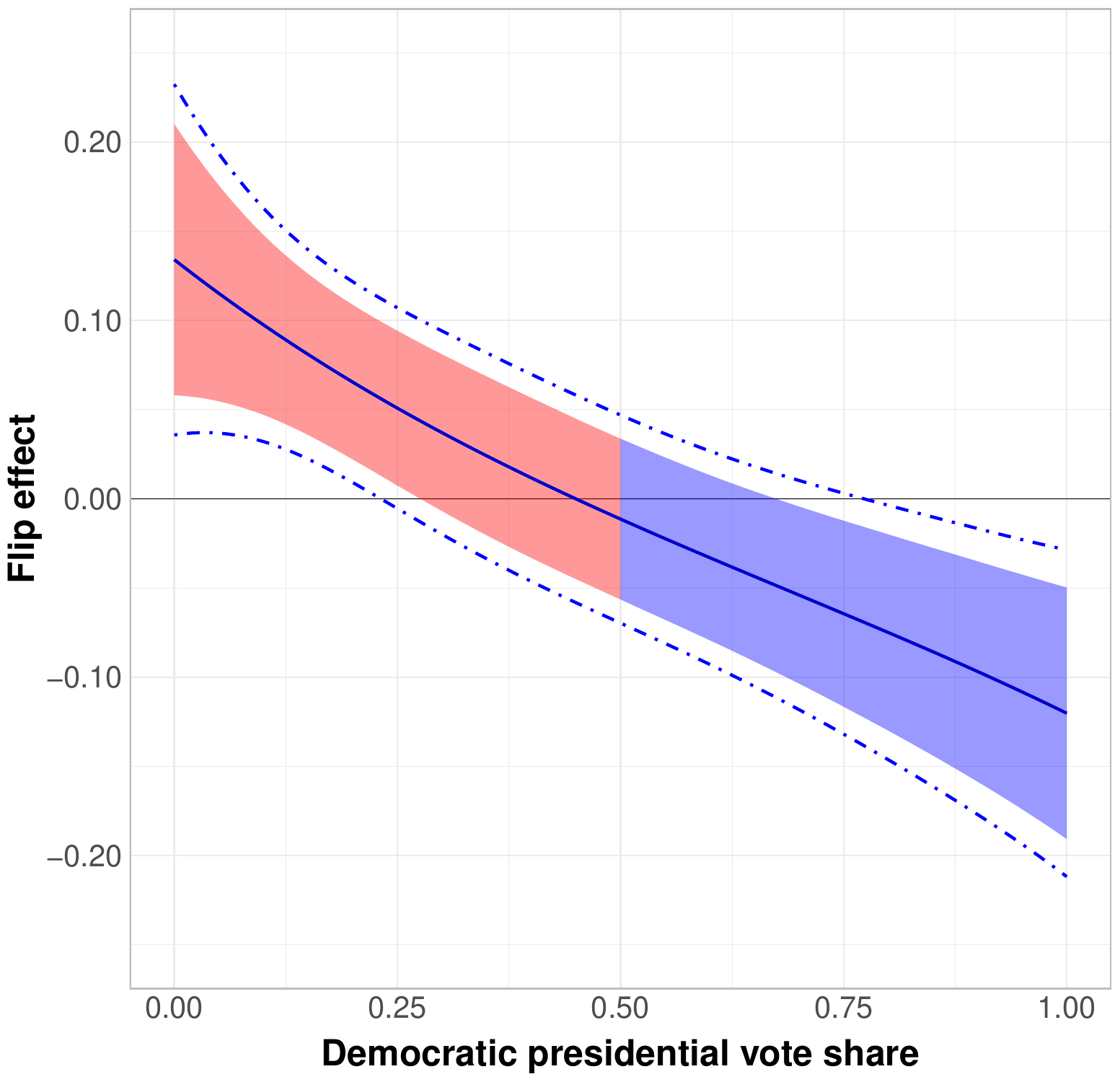}\hfill
\includegraphics[width=\sidebysidepicturewidht\linewidth]{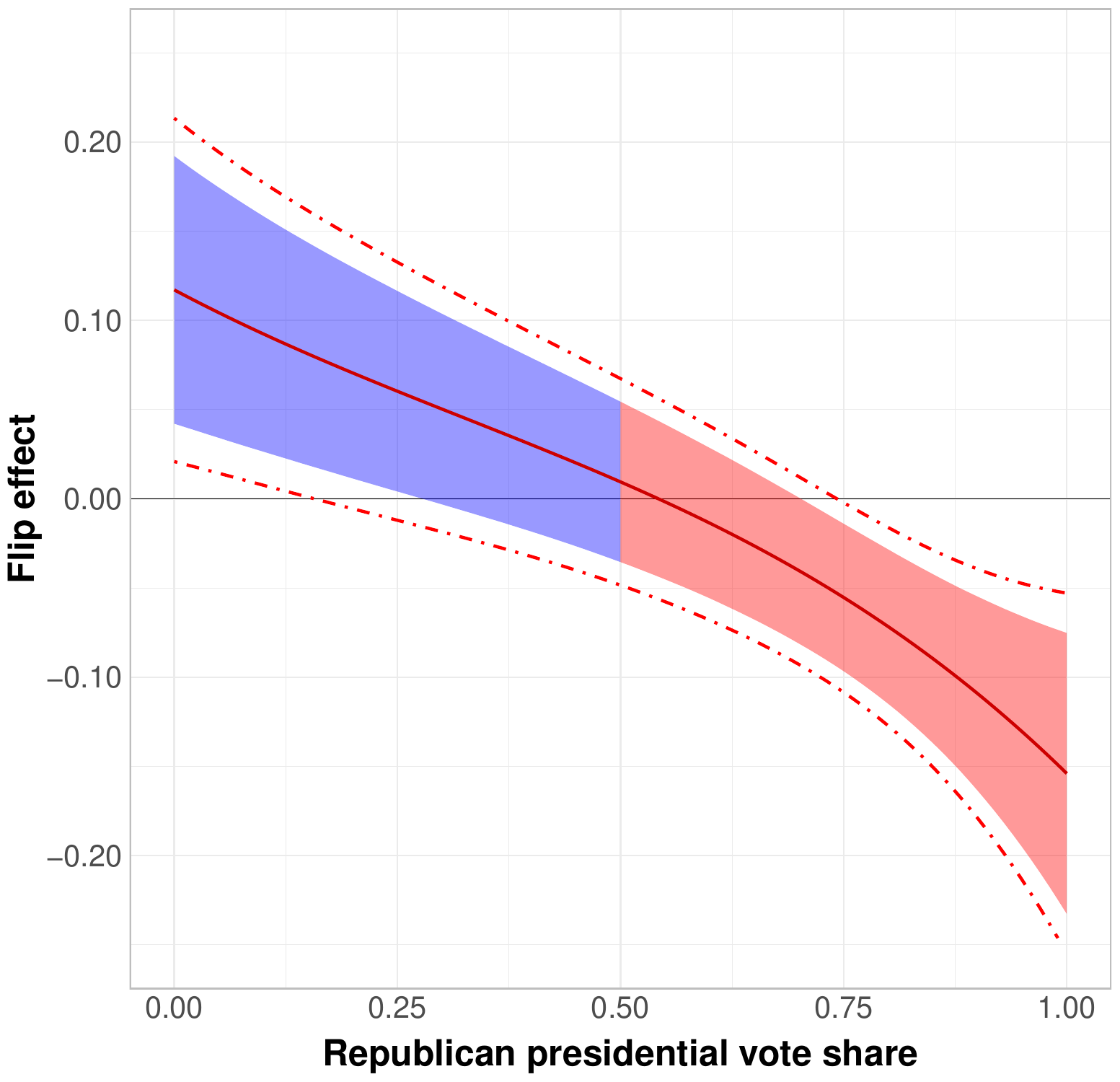}}

\caption*{\emph{Note.}
The figure presents the estimated flip effect for judicial candidates in contests \emph{without} party labels (cubic specification, $q = 3$).
The left panel shows the estimated flip effect, $\hat f_d(x)$,
for Democratic judicial candidates as a function of $x$, the vote share of the Democratic presidential candidate in the same election and precinct (solid blue line).
Shaded region: 95\% pointwise confidence intervals.
Outer blue dashed lines: 95\% uniform confidence band.
The right panel shows the analogous flip effect, $\hat f_r(x)$,
for Republican judicial candidates (lines in red).}
\end{adjustwidth}

\vspace{0.5em}

\captionof{table}{\textbf{Estimation results for flip effect without party labels (cubic specification)}}
\label{tab:nonpartisan-cubic-detailed}
\centering
\begin{adjustwidth}{-1cm}{-1cm}
\scalebox{0.95}{%
\footnotesize
\begin{tabular}{c@{\hskip 16pt}
                r@{\hskip 8pt}
                c@{\hskip 8pt}
                r@{\hskip 8pt}
                c@{\hskip 8pt}
                r@{\hskip 0pt}
                r@{\hskip 0pt}
                r@{\hskip 0.166667em}
                r@{\hskip 0pt}
                l@{\hskip 0pt} 
                c@{\hskip 16pt}
                r@{\hskip 8pt}
                c@{\hskip 8pt}
                r@{\hskip 8pt}
                c@{\hskip 8pt}
                r@{\hskip 0pt}
                r@{\hskip 0pt}
                r@{\hskip 0.166667em}
                r@{\hskip 0pt}
                l}

\toprule
&
\multicolumn{9}{c}{\textbf{Democratic party}}
&
&
\multicolumn{9}{c}{\textbf{Republican party}} \\
\cmidrule(lr){2-10}
\cmidrule(lr){12-20}

\multicolumn{3}{l}{Partisan-voting mistakes ($\hat \tau$)}
&
\multicolumn{7}{r}{{\DEMNonpartisanCubicVotingMistakesESTPercentage\,
{(SE: \DEMNonpartisanCubicVotingMistakesSEPercentage)}}}
&
&
\multicolumn{9}{r}{{\REPNonpartisanCubicVotingMistakesESTPercentage\,
{(SE: \REPNonpartisanCubicVotingMistakesSEPercentage)}}}
\\

\multicolumn{3}{l}{Test of zero flip effect}
&
\multicolumn{7}{r}{\DEMNonpartisanCubicTestZeroFlipDecision,
p-value: \DEMNonpartisanCubicTestZeroFlipP}
&
&
\multicolumn{9}{r}{\REPNonpartisanCubicTestZeroFlipDecision,
p-value: \REPNonpartisanCubicTestZeroFlipP}
\\

\multicolumn{3}{l}{Test of homogeneous flip effect}
&
\multicolumn{7}{r}{\DEMNonpartisanCubicTestHomogeneousFlipDecision,
p-value: \DEMNonpartisanCubicTestHomogeneousFlipP}
&
&
\multicolumn{9}{r}{\REPNonpartisanCubicTestHomogeneousFlipDecision,
p-value: \REPNonpartisanCubicTestHomogeneousFlipP}
\\

\midrule
\makecell[c]{\small\textit{Coefficient}}
  & \makecell[c]{\small\textit{Estimate}}
  & \makecell[c]{\small\textit{SE}}
  & \makecell[c]{\small\textit{t}}
  & \makecell[c]{\small\textit{p-value}}
  & \multicolumn{5}{c}{\makecell[c]{\small\emph{95\% CI}}}
  &
  & \makecell[c]{\small\textit{Estimate}}
  & \makecell[c]{\small\textit{SE}}
  & \makecell[c]{\small\textit{t}}
  & \makecell[c]{\small\textit{p-value}}
  & \multicolumn{5}{c}{\makecell[c]{\small\emph{95\% CI}}} \\
\midrule

$\hat\theta_{0}$
  & \DEMNonpartisanCubicThetaZeroEST
  & \DEMNonpartisanCubicThetaZeroSE
  & \DEMNonpartisanCubicThetaZeroT
  & \DEMNonpartisanCubicThetaZeroP
  & [
  & \DEMNonpartisanCubicThetaZeroCILeft
  & ,
  & \DEMNonpartisanCubicThetaZeroCIRight
  & ]
  &
  & \REPNonpartisanCubicThetaZeroEST
  & \REPNonpartisanCubicThetaZeroSE
  & \REPNonpartisanCubicThetaZeroT
  & \REPNonpartisanCubicThetaZeroP
  & [
  & \REPNonpartisanCubicThetaZeroCILeft
  & ,
  & \REPNonpartisanCubicThetaZeroCIRight
  & ]
  \\

$\hat\theta_{1}$
  & \DEMNonpartisanCubicThetaOneEST
  & \DEMNonpartisanCubicThetaOneSE
  & \DEMNonpartisanCubicThetaOneT
  & \DEMNonpartisanCubicThetaOneP
  & [
  & \DEMNonpartisanCubicThetaOneCILeft
  & ,
  & \DEMNonpartisanCubicThetaOneCIRight
  & ]
  &
  & \REPNonpartisanCubicThetaOneEST
  & \REPNonpartisanCubicThetaOneSE
  & \REPNonpartisanCubicThetaOneT
  & \REPNonpartisanCubicThetaOneP
  & [
  & \REPNonpartisanCubicThetaOneCILeft
  & ,
  & \REPNonpartisanCubicThetaOneCIRight
  & ]
  \\

$\hat\theta_{2}$
  & \DEMNonpartisanCubicThetaTwoEST
  & \DEMNonpartisanCubicThetaTwoSE
  & \DEMNonpartisanCubicThetaTwoT
  & \DEMNonpartisanCubicThetaTwoP
  & [
  & \DEMNonpartisanCubicThetaTwoCILeft
  & ,
  & \DEMNonpartisanCubicThetaTwoCIRight
  & ]
  &
  & \REPNonpartisanCubicThetaTwoEST
  & \REPNonpartisanCubicThetaTwoSE
  & \REPNonpartisanCubicThetaTwoT
  & \REPNonpartisanCubicThetaTwoP
  & [
  & \REPNonpartisanCubicThetaTwoCILeft
  & ,
  & \REPNonpartisanCubicThetaTwoCIRight
  & ]
  \\

$\hat\theta_{3}$
  & \DEMNonpartisanCubicThetaThreeEST
  & \DEMNonpartisanCubicThetaThreeSE
  & \DEMNonpartisanCubicThetaThreeT
  & \DEMNonpartisanCubicThetaThreeP
  & [
  & \DEMNonpartisanCubicThetaThreeCILeft
  & ,
  & \DEMNonpartisanCubicThetaThreeCIRight
  & ]
  &
  & \REPNonpartisanCubicThetaThreeEST
  & \REPNonpartisanCubicThetaThreeSE
  & \REPNonpartisanCubicThetaThreeT
  & \REPNonpartisanCubicThetaThreeP
  & [
  & \REPNonpartisanCubicThetaThreeCILeft
  & ,
  & \REPNonpartisanCubicThetaThreeCIRight
  & ]
  \\

\bottomrule
\end{tabular}%
}

\caption*{\emph{Note.}
The table presents estimates of the flip effect for judicial candidates in contests \emph{without} party labels (cubic specification, $q = 3$).
The table reports estimates of partisan-voting mistakes \eqref{eq:partisan-voting-mistakes}
and results for the test of zero flip effect \eqref{test:ZeroFlipEffect}
and the test of homogeneous flip effect \eqref{test:HomogeneousFlipEffect}.
These estimates are based on {\nNonpartisanObservationsOneParty} precinct-level observations from Democratic judicial candidates across {\nNonpartisanContests} races \emph{without} party labels.
Of these races, {\nNonpartisanContestsFlipped} exhibit a party-order flip relative to the presidential race,
while the remaining {\nNonpartisanContestsNotFlipped} do not.
The Republican estimates are based on observations from Republican judicial candidates in the same set of races.
}
\end{adjustwidth}
\end{figure}
}

In this section, we consider a \emph{cubic specification} of the flip effect by setting $q = 3$ in \eqref{eq:f-qpoly-specification}.
We then estimate the flip effect in \eqref{eq:outcome-model} and the 
partisan-voting mistakes in \eqref{eq:partisan-voting-mistakes} separately 
for the Democratic and Republican parties, denoting these estimates by 
$\hat f_d$, $\hat f_r$, $\hat \tau_d$, and $\hat \tau_r$, respectively.
This specification is well-suited to our context, as it allows the flip effect to be either monotonic or non-monotonic, and either antisymmetric or asymmetric about $x=0.5$.\footnote{A function $f: \mathbb{R} \rightarrow \mathbb{R}$ is antisymmetric about $x$ if $f(x + h) = - f(x - h)$ for all $h \in \mathbb{R}$.}
Other alternatives---such as constant, linear, or quadratic---would impose monotonicity, antisymmetry, or both by construction (see Appendix~\ref{se:linear&constant-specifications}).
We estimate the flip effect and the partisan-voting mistakes using {\nNonpartisanObservationsOneParty} precinct-level observations across {\nNonpartisanContestsFlipped} flipped and {\nNonpartisanContestsNotFlipped} non-flipped NC statewide judicial races \emph{without} party designations (see Section \ref{se:data}).
Table~\ref{tab:nonpartisan-cubic-detailed} and Figure~\ref{fig:AverageTreatmentEffectNoPartyLabels} summarize our findings.

We estimate that $\hat \tau_d = \DEMNonpartisanCubicVotingMistakesESTPercentage$ (95\% confidence interval: [\DEMNonpartisanCubicVotingMistakesCILeftPercentage, \DEMNonpartisanCubicVotingMistakesCIRightPercentage]) of Democratic partisan voters and 
$\hat \tau_r = \REPNonpartisanCubicVotingMistakesESTPercentage$ (95\% confidence interval: [\REPNonpartisanCubicVotingMistakesCILeftPercentage, \REPNonpartisanCubicVotingMistakesCIRightPercentage]) of Republican partisan voters cast their votes incorrectly due to party-order flips.

Figure \ref{fig:AverageTreatmentEffectNoPartyLabels} plots the estimated flip effect under the cubic specification for judicial candidates without party designations, as a function of the vote share of
the corresponding presidential candidate of the same party.
In both panels of Figure \ref{fig:AverageTreatmentEffectNoPartyLabels}, the solid lines represent the estimates of the conditional average effect of a party-order flip as a function of the presidential vote share $x$, the shaded regions provide 95\% cluster-robust
pointwise confidence intervals, and the outer dashed lines correspond to the 95\% uniform confidence band.

The left panel of Figure \ref{fig:AverageTreatmentEffectNoPartyLabels} depicts the flip effect for Democratic judicial candidates.
The $x$-axis plots the vote share of the Democratic presidential candidate, and the $y$-axis plots the estimated conditional average flip effect $\hat f_d (x) = \hat \theta_{d0} + \hat \theta_{d1} x + \hat \theta_{d2} x^2 + \hat \theta_{d3} x^3$ (solid blue line), where the coefficient estimates $\hat \theta_{d0}, \hat\theta_{d1}, \hat\theta_{d2}, \hat\theta_{d3}$ are reported in Table \ref{tab:nonpartisan-cubic-detailed}. 
For a given Democratic presidential vote share $x$, the value $\hat f_d(x)$ estimates the  causal impact of a party-order flip on the vote share of a Democratic judicial candidate.
\emph{Low} vote shares (small $x$) for the Democratic presidential candidate correspond to a \emph{positive} conditional average flip effect 
($\hat f_d (x) > 0)$, 
while \emph{high} presidential vote shares (large $x$) correspond to a \emph{negative} conditional average flip effect ($\hat f_d ( x ) < 0)$.
In other words, in a precinct with a smaller Democratic base, the effect of a party-order flip and the absence of party labels help the Democratic judicial candidate whose vote share increases, on average, because the larger share of votes cast incorrectly is by Republican partisan voters.
In contrast, in a precinct with a larger Democratic base, the biggest share of incorrectly cast votes comes from Democratic partisan voters  thereby reducing the vote share of the Democratic judicial candidate on average.

In the right panel of Figure \ref{fig:AverageTreatmentEffectNoPartyLabels},
we find the estimated conditional average flip effect under the cubic specification for Republican judicial candidates, $\hat f_r(x) = \hat\theta_{r0} + \hat\theta_{r1} x + \hat\theta_{r2} x^2 + \hat\theta_{r3} x^3$, 
and obtain similar findings (see also Table \ref{tab:nonpartisan-cubic-detailed}). 
All of the earlier observations about the left panel apply here as well with the caveat that, in the right panel, small values of $x$ correspond to precincts in which the Republican presidential candidate has few votes, while large values of $x$ correspond to precincts in which the Republican presidential candidate receives a large fraction of votes. 

Our estimates also allow us to test the structure of the flip effect 
for each party using joint Wald tests with cluster-robust variance estimators. 
We \DEMNonpartisanCubicTestZeroFlipDecision{} the null hypothesis of a zero flip 
effect \eqref{test:ZeroFlipEffect} for both parties (p-value: 
$\DEMNonpartisanCubicTestZeroFlipP$ for Democratic candidates, 
$\REPNonpartisanCubicTestZeroFlipP$ for Republican candidates), 
and we \DEMNonpartisanCubicTestHomogeneousFlipDecision{} the null hypothesis of a 
homogeneous flip effect \eqref{test:HomogeneousFlipEffect} (p-value: 
\DEMNonpartisanCubicTestHomogeneousFlipP{} for Democrats, 
\REPNonpartisanCubicTestHomogeneousFlipP{} for Republicans).
The results of the latter test are particularly noteworthy because they reject a homogeneous flip effect for both parties, supporting the heterogeneous specification that is central to our analysis.

Our analysis also allows us to test whether partisan-voting mistakes vary by party.
We find that $\hat \tau_d - \hat \tau_r = \hat f_r(1)- \hat f_d(1)$ is not statistically different from zero. 
We can also test if partisan-voting mistakes cancel out when the presidential vote share is evenly split. 
Our analysis shows that both $\hat f_r(1/2)$ and $\hat f_d(1/2)$ are not statistically significant.
Taken together, these results support the homogeneity of partisan-voting mistakes across parties.

We note here that, although our estimation procedure imposes no symmetry or other cross-party constraints linking the Democratic and Republican flip effects (Remark~\ref{rem:parti-specific-estimation}), the estimated effects are broadly consistent with mirror-image behavior. 
In particular, the estimated flip effects for Democratic and Republican candidates have similar magnitudes and opposite signs across the range of presidential vote shares as shown in Figure~\ref{fig:AverageTreatmentEffectNoPartyLabels}
and Table~\ref{tab:nonpartisan-cubic-detailed}.
While a constraint such as $f_d(x) = -f_r(1-x)$ could be motivated by a stylized two-party setting, we do not impose it because voting behavior may differ systematically across parties and because Democratic and Republican presidential vote shares do not generally sum to one due to third-party candidates and write-ins.

Finally, we close this section by emphasizing that heterogeneity
with respect to presidential vote share is essential for accurately estimating the causal effect of a party-order flip.
As we demonstrate in Appendix~\ref{se:linear&constant-specifications}, 
an analysis limited to the average treatment effect (ATE) 
would misleadingly suggest that
party-order flips 
are inconsequential.
In our case, this misinterpretation arises from the roughly balanced partisan composition of the North Carolina electorate, combined with the reciprocal nature of voting mistakes, where support lost by one party’s candidate is transferred to the opposing party’s candidate.
By accounting for heterogeneity, however, our model estimates that average proportion of partisan-voting mistakes of the Democratic and the Republican party are respectively $\DEMNonpartisanCubicVotingMistakesESTPercentage$ 
and $\REPNonpartisanCubicVotingMistakesESTPercentage$. 
As a point of comparison, we note that in the 2000 U.S. Presidential Election less than 1\% of Democratic voters in Florida's Palm Beach County were misled by the infamous butterfly ballot \citep{SinclairMarkMooreLavisSoldat:Nature2000electoral,WandShottsSekhon:AmPoliSciRev2001,Smith:StaSci2002}.

\subsection{Placebo test: conditional average flip effect with party labels}\label{se:results-yes-party-labels}

\begin{figure}[t]     
\linespread{1.25}\selectfont
\begin{adjustwidth}{-1cm}{-1cm}
 \caption{\textbf{Placebo test: flip effect with party labels (cubic specification)}}
  \label{fig:AverageTreatmentEffectWPartyLabels}
 {\includegraphics[width=\sidebysidepicturewidht\linewidth]{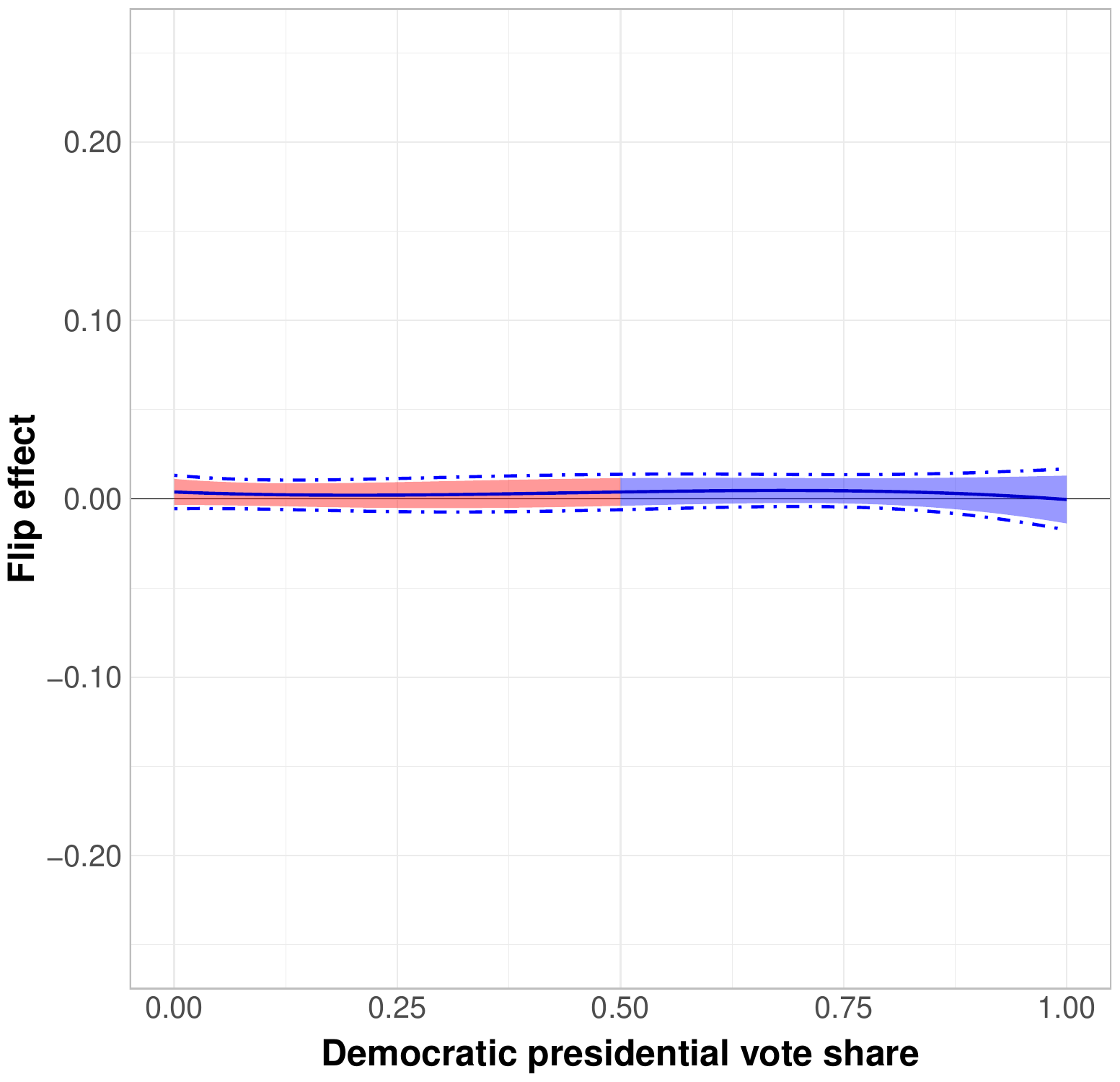}\hfill
 \includegraphics[width=\sidebysidepicturewidht\linewidth]{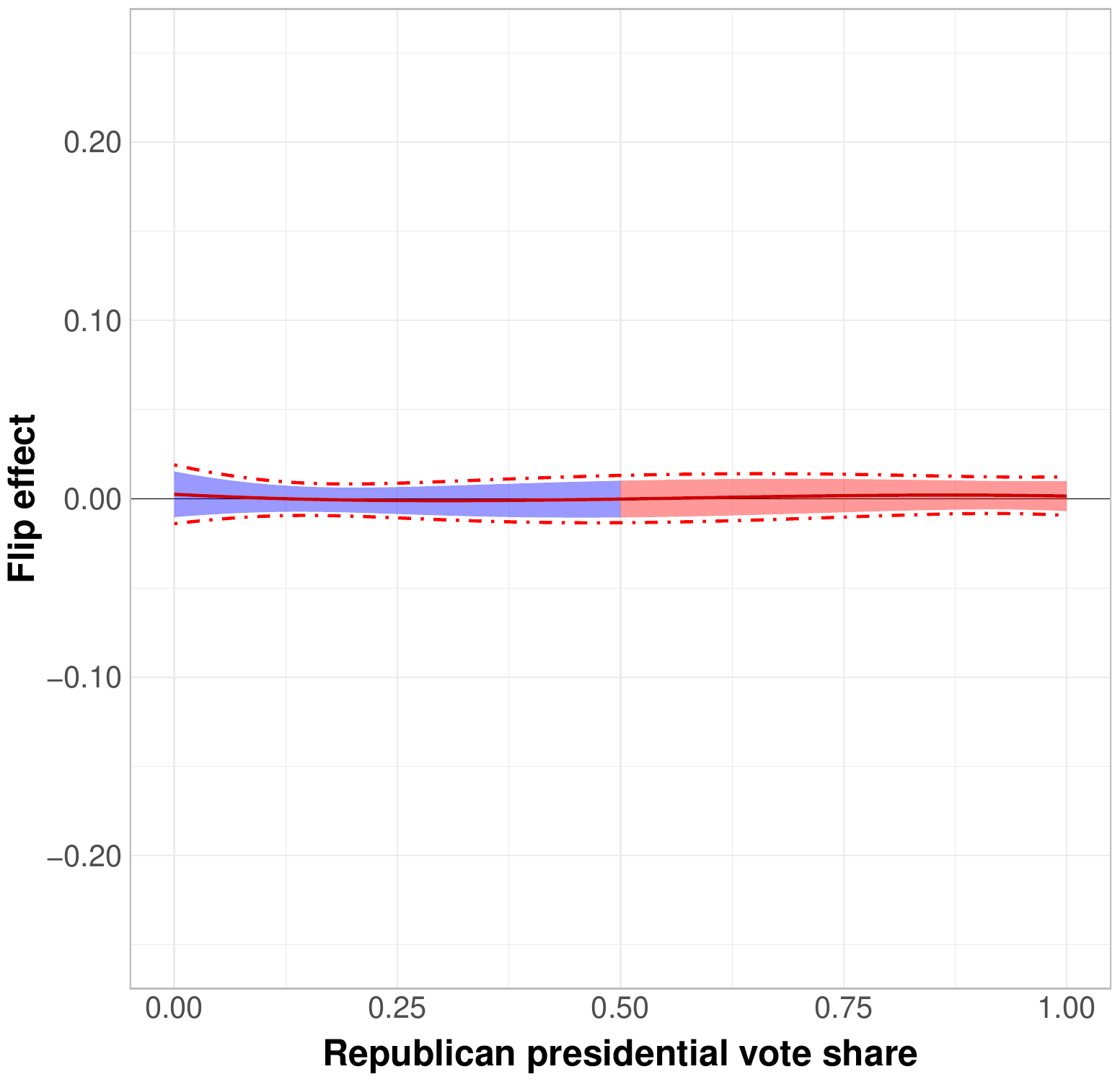}}
 \caption*{\emph{Note.}
 The left panel shows the estimated placebo-test flip effect, $\hat f_d (x)$,
 for Democratic judicial candidates with party designation as a function of $x$, the vote share of the Democratic presidential candidate in the same election and precinct (solid blue line). 95\% pointwise confidence intervals are given by the shaded region, while uniform confidence intervals are depicted by the outer blue dashed lines. The right panel shows the analogous placebo-test flip effect, $\hat f_r (x)$,
 for Republican judicial candidates (lines in red).}
 \end{adjustwidth}
\end{figure}

To support the hypothesized cause-and-effect relationship, 
we leverage the statutory change that required party affiliations to be printed on 
the ballot for judicial contests starting with the 2016 Court of Appeals races and 
extending to all statewide judicial races from 2020 onward 
(see Section~\ref{se:arrangement-of-ballots}).
With this statutory change, quasi-randomized party-order flips relative to the 
presidential race continue to occur, so the variation in treatment status is still present.
What has changed is the information provided to voters, because party affiliation is now 
directly observable on the ballot, voters no longer need to rely on party order as a 
cue to infer which candidate belongs to which party.
Our hypothesized mechanism for the flip effect is precisely that voters \emph{use 
party order as a cue for party affiliation in the absence of party labels}.
Therefore, if this mechanism is correct, the flip effect should disappear when party 
labels are present, even though party-order flips still occur and all other features 
of the election (ballot order, candidate gender, incumbency, contest-level covariates) 
remain unchanged.
This design rules out alternative explanations,
thereby isolating the informational mechanism at the core of our model.

Our placebo sample consists of {\nPartisanObservationsOneParty} precinct-level observations from {\nPartisanContestsFlipped} flipped and {\nPartisanContestsNotFlipped} non-flipped North Carolina statewide judicial races with party designations printed on the ballot. 
The placebo analysis mirrors Section~\ref{se:results-no-party-labels},
employing the same cubic specification.
Figure~\ref{fig:AverageTreatmentEffectWPartyLabels}---the analog of 
Figure~\ref{fig:AverageTreatmentEffectNoPartyLabels} for judicial contests with party 
labels---shows that the flip effect is neutralized. 
Estimated effects are close to zero across the full range of presidential vote shares and not statistically 
significant for either party (see Table~\ref{tab:partisan-cubic-detailed} in 
Appendix~\ref{app:placebo}).


\section{Conclusion}\label{se:conclusions}

We propose a causal inference approach to identify how treatment and information (whether displayed, concealed, implicit, or explicit) affect agents' behavior in choice settings.
A key feature of our approach is its ability to exploit heterogeneity to estimate not only the net effect of the treatment, but also its offsetting components.

We demonstrate this approach in the context of electoral choices,
where ballot design can unintentionally distort voter behavior. 
Specifically, we show that ballots mixing contests with and without party labels 
can mislead a significant number of voters.
Partisan voters may incorrectly infer the party affiliation of candidates in races without party labels by referencing the (party) order of the candidates in labeled races. 
As a result, outcomes in races without party labels 
may not reflect the true preferences of the electorate.
To accurately capture voter intent, such ballot designs should be avoided.

Our empirical analysis leverages changes in the North Carolina General Statute that, 
over the past 20 years, affected the arrangement and order of candidates on official ballots.
These changes create a quasi-randomized natural experiment with contest-level treatment assignment 
in NC judicial elections, allowing us to identify both the net flip effect and the total partisan-voting mistakes. 
Using precinct-level election and demographic data, we leverage DML to estimate the heterogeneous effect of a party-order flip for voters of each party.
We find that 
{\DEMNonpartisanCubicVotingMistakesESTPercentage} 
(95\% confidence interval: [\DEMNonpartisanCubicVotingMistakesCILeftPercentage, \DEMNonpartisanCubicVotingMistakesCIRightPercentage]) 
of Democratic partisan voters and 
{\REPNonpartisanCubicVotingMistakesESTPercentage}
(95\% confidence interval: [\REPNonpartisanCubicVotingMistakesCILeftPercentage, \REPNonpartisanCubicVotingMistakesCIRightPercentage]) 
of Republican partisan voters cast their votes incorrectly due to party-order flips.






\newpage

\appendix

\section{Variables}\label{appendix:variables}

The variables listed below are included in four separate datasets: 
one for the main analysis pertaining to judicial races without party 
designations for candidates supported by the Democratic party, one for the main analysis 
pertaining to judicial races without party designations for candidates supported by the Republican party, and the corresponding two datasets for the placebo analysis 
pertaining to judicial races with party designations. In each dataset, 
all vote shares and candidate-level covariates refer to candidates 
supported by the same party as the judicial candidate.

\subsection*{Judicial Race Characteristics}
\begin{itemize}
    \item \texttt{candidate.vote.share}: precinct-level vote share of the judicial candidate (dependent variable)
    \item \texttt{year}: election year fixed effect
    \item \texttt{contest.type}: type of judicial race (Supreme Court, Court of Appeals)
    \item \texttt{contest.position.on.ballot}: position of the judicial contest on the ballot
    \item \texttt{candidate.position.in.contest}: candidate's position within the judicial contest
    \item \texttt{candidate.gender}: gender of the judicial candidate
    \item \texttt{candidate.incumbent}: indicator for whether the judicial candidate is an incumbent
    \item \texttt{candidate.incumbent.and.party.identifier}: interaction of incumbency status and party identifier
\end{itemize}

\subsection*{Treatment Variable}
\begin{itemize}
    \item \texttt{flip}: a contest-level indicator variable equal to one if the party order 
    of the judicial race differs from that of the U.S.\ presidential race in the same 
    election, and equal to zero otherwise
\end{itemize}

\subsection*{Concurrent Statewide Race Covariates}

For each concurrent statewide race, the covariates include the precinct-level vote share of the candidate supported by the same party as the judicial candidate and an indicator for whether that candidate carried the precinct. 
The list below does not include covariates pertaining to the State Attorney General race. 
Although this election takes place in presidential years alongside the other statewide executive offices, the corresponding covariates were excluded from the analysis because the 2012 race was uncontested.

\begin{itemize}
    \item \texttt{potus\_candidate.vote.share}: presidential candidate vote share
    \item \texttt{potus\_candidate.winning.in.precinct}: indicator for whether the presidential candidate carried the precinct
    \item \texttt{governor\_candidate.vote.share}: gubernatorial candidate vote share
    \item \texttt{governor\_candidate.winning.in.precinct}: indicator for whether the gubernatorial candidate carried the precinct
    \item \texttt{ltgovernor\_candidate.vote.share}: Lieutenant Governor candidate vote share
    \item \texttt{ltgovernor\_candidate.winning.in.precinct}: indicator for whether the Lieutenant Governor candidate carried the precinct
    \item \texttt{auditor\_candidate.vote.share}: Auditor candidate vote share
    \item \texttt{auditor\_candidate.winning.in.precinct}: indicator for whether the Auditor candidate carried the precinct
    \item \texttt{comm.agriculture\_candidate.vote.share}: Commissioner of Agriculture candidate vote share
    \item \texttt{comm.agriculture\_candidate.winning.in.precinct}: indicator for whether the Commissioner of Agriculture candidate carried the precinct
    \item \texttt{comm.insurance\_candidate.vote.share}: Commissioner of Insurance candidate vote share
    \item \texttt{comm.insurance\_candidate.winning.in.precinct}: indicator for whether the Commissioner of Insurance candidate carried the precinct
    \item \texttt{comm.labor\_candidate.vote.share}: Commissioner of Labor candidate vote share
    \item \texttt{comm.labor\_candidate.winning.in.precinct}: indicator for whether the Commissioner of Labor candidate carried the precinct
    \item \texttt{sec.state\_candidate.vote.share}: Secretary of State candidate vote share
    \item \texttt{sec.state\_candidate.winning.in.precinct}: indicator for whether the Secretary of State candidate carried the precinct
    \item \texttt{superintendent\_candidate.vote.share}: Superintendent of Public Instruction candidate vote share
    \item \texttt{superintendent\_candidate.winning.in.precinct}: indicator for whether the Superintendent of Public Instruction candidate carried the precinct
    \item \texttt{treasurer\_candidate.vote.share}: Treasurer candidate vote share
    \item \texttt{treasurer\_candidate.winning.in.precinct}: indicator for whether the Treasurer candidate carried the precinct
\end{itemize}

\subsection*{Actual Voter Demographic Covariates}

Variables are named \texttt{A.GPCESV.pct} and represent the precinct-level proportion of actual voters belonging to a given demographic group, where \texttt{GPCESV} denotes the demographic group defined by the cross-tabulation of the six features \texttt{G} (age group), \texttt{P} (party registration), \texttt{C} (race), \texttt{E} (ethnicity), \texttt{S} (sex), and \texttt{V} (voting method).
A value of \texttt{0} in any 
position denotes the marginal across all groups in that feature. Only 
variables whose precinct-level standard deviation falls in the top {\pctCovariates} across all groups are retained for the analysis.

\begin{itemize}
    \item[\texttt{G:}] \textbf{Age group.}
        \texttt{0}: any;
        \texttt{1}: less than 18 or invalid;
        \texttt{2}: 18 to 25;
        \texttt{3}: 26 to 40;
        \texttt{4}: 41 to 65;
        \texttt{5}: over 65.
    \item[\texttt{P:}] \textbf{Party registration.}
        \texttt{0}: any;
        \texttt{1}: Democrat;
        \texttt{2}: Libertarian;
        \texttt{3}: Republican;
        \texttt{4}: Unaffiliated;
        \texttt{5}: other.
    \item[\texttt{C:}] \textbf{Race.}
        \texttt{0}: any;
        \texttt{1}: Asian;
        \texttt{2}: Black;
        \texttt{3}: American Indian;
        \texttt{4}: Multiracial;
        \texttt{5}: Other;
        \texttt{6}: Unknown;
        \texttt{7}: White.
    \item[\texttt{E:}] \textbf{Ethnicity.}
        \texttt{0}: any;
        \texttt{1}: Hispanic;
        \texttt{2}: Non-Hispanic;
        \texttt{3}: Unknown.
    \item[\texttt{S:}] \textbf{Sex.}
        \texttt{0}: any;
        \texttt{1}: Female;
        \texttt{2}: Male;
        \texttt{3}: Unknown.
    \item[\texttt{V:}] \textbf{Voting method.}
        \texttt{0}: any;
        \texttt{1}: absentee;
        \texttt{2}: curbside;
        \texttt{3}: in-person;
        \texttt{4}: provisional;
        \texttt{5}: transfer;
        \texttt{6}: legacy;
        \texttt{7}: early voting.
\end{itemize}

\subsubsection*{Registered Voter Demographic Covariates}
Variables are named \texttt{R.GPCES0.pct} and represent the precinct-level proportion of registered voters belonging to a given demographic group.
As above, \texttt{GPCES0} identifies the demographic group through the cross-tabulation of the five features \texttt{G}, \texttt{P}, \texttt{C}, \texttt{E}, and \texttt{S}. The sixth position is fixed at \texttt{0} because voting method is not observed for registered voters.
Only variables whose precinct-level standard deviation 
falls in the top {\pctCovariates} across all groups are retained for the analysis.

\subsection*{Turnout Demographic Covariates}
Variables are named \texttt{T.GPCES0.pct} and represent the precinct-level 
proportion of registered voters in a given demographic group who cast a 
ballot.
As above, \texttt{GPCES0} identifies the demographic group through the cross-tabulation of the five features \texttt{G}, \texttt{P}, \texttt{C}, \texttt{E}, and \texttt{S}.
The sixth position is always \texttt{0} as voting method is not applicable for 
turnout rates. Only variables whose precinct-level standard deviation falls 
in the top {\pctCovariates} across all groups are retained for the analysis.

\section{Alternative tests}\label{appendix:AlternativeTests}

In Section \ref{se:causal-inference-framework}, 
we introduce the test of zero flip effect and the test of homogeneous flip effect,
both of which, under the polynomial specification in \eqref{eq:f-qpoly-specification},
are implemented as Wald tests on restrictions of the polynomial coefficients.

More generally, however, the null hypothesis for the test of zero flip effect is:
$$
H^z_0: f(x) = 0 \text{ for all } x \in [0,1].
$$
Leveraging the approximate linear representation of the DML estimators, this hypothesis can be tested using the statistic 
$$ 
T_n = \sup_{x \in [0,1] } \left| \frac{\hat f(x)}{\hat\sigma(\hat f(x))}\right|,
$$
where $\hat\sigma(\hat f(x))$ is an estimate of the pointwise standard error of $\hat f(x)$. 
Because of the supremum, the distribution of $T_n$ under the null hypothesis is non-Gaussian,
but it  can be approximated via a bootstrap procedure. 
Using this procedure, we {\DEMNonpartisanCubicTestZeroFlipDecision} the null hypothesis of zero flip effect, with p-value {\DEMNonpartisanCubicTestZeroFlipBootstrapP} for the Democratic party and {\REPNonpartisanCubicTestZeroFlipBootstrapP} for the Republican party.
This conclusion is consistent with the findings in Section \ref{se:results-no-party-labels}, 
where Wald tests applied under the polynomial specification also {\DEMNonpartisanCubicTestZeroFlipDecision} the null of zero flip effect.

Similarly, for the test of homogeneous flip effect, the null hypothesis is given by 
$$
H^h_0: \text{ there exists $\cbar$ such that } f(x) = \cbar  
\text{ for all } x \in [0,1],
$$
which can be tested using the statistic
$$ 
H_n = \inf_{c \in \mathbb{R}} \left\{\sup_{x \in [0,1] } \left| \frac{\hat f(x) - c}{\hat\sigma(\hat f(x))}\right|\right\}.
$$
The distribution of this test statistic under the null is also non-Gaussian but can be approximated using a bootstrap procedure. 
Based on this approach, we {\DEMNonpartisanCubicTestHomogeneousFlipDecision} the null hypothesis of homogeneous flip effect, with p-value of {\DEMNonpartisanCubicTestHomogeneousFlipBootstrapP} for the Democratic party and {\REPNonpartisanCubicTestHomogeneousFlipBootstrapP} for the Republican party.
This result aligns with the conclusions from Section \ref{se:results-no-party-labels}, where we also {\DEMNonpartisanCubicTestHomogeneousFlipDecision} the null of a homogeneous flip effect using the polynomial specification \eqref{eq:f-qpoly-specification} and Wald tests.

\section{Flip effect under linear and constant specifications}\label{se:linear&constant-specifications}

\begin{table}[t!]
\linespread{1.25}\selectfont

\begin{adjustwidth}{-1cm}{-1cm}

\caption{\textbf{Flip effect without party labels under cubic, linear, and constant specifications}}
\label{tab:nonpartisan-threespecifications-summary}
\centering
\scalebox{1.0}{
\footnotesize
\begin{tabular}{r|
                 r@{\hskip 0pt}l 
                 r@{\hskip 0pt}l 
                 r@{\hskip 0pt}l|
                 r@{\hskip 0pt}l  
                 r@{\hskip 0pt}l  
                 r@{\hskip 0pt}l}
\toprule
& \multicolumn{6}{c|}{\textbf{Democratic party}} 
& \multicolumn{6}{c}{\textbf{Republican party}} \\
\cmidrule(r){2-7} \cmidrule(l){8-13}
\textit{Flip effect specification} 
  & \multicolumn{2}{c}{\itshape Cubic} 
  & \multicolumn{2}{c}{\itshape Linear} 
  & \multicolumn{2}{c|}{\itshape Constant} 
  & \multicolumn{2}{c}{\itshape Cubic} 
  & \multicolumn{2}{c}{\itshape Linear} 
  & \multicolumn{2}{c}{\itshape Constant} \\
\midrule

Partisan-voting mistakes 
  & \DEMNonpartisanCubicVotingMistakesESTPercentage  
  & \significancemarker{\DEMNonpartisanCubicVotingMistakesPPrecision}
  & \DEMNonpartisanLinearVotingMistakesESTPercentage  
  & \significancemarker{\DEMNonpartisanLinearVotingMistakesPPrecision} 
  &  
  &  
  & \REPNonpartisanCubicVotingMistakesESTPercentage  
  & \significancemarker{\REPNonpartisanCubicVotingMistakesPPrecision}  
  & \REPNonpartisanLinearVotingMistakesESTPercentage  
  & \significancemarker{\REPNonpartisanLinearVotingMistakesPPrecision}  
  & 
  & 
  \\ 
& {(\DEMNonpartisanCubicVotingMistakesSEPercentage)}  
  &  
  & {(\DEMNonpartisanLinearVotingMistakesSEPercentage)}  
  &  
  &  
  &  
  & {(\REPNonpartisanCubicVotingMistakesSEPercentage)}  
  &  
  & {(\REPNonpartisanLinearVotingMistakesSEPercentage)}  
  &
  &  
  &  
  \\

Test of zero flip effect 
  & \DEMNonpartisanCubicTestZeroFlipDecision  
  & \significancemarker{\DEMNonpartisanCubicTestZeroFlipPPrecision}  
  & \DEMNonpartisanLinearTestZeroFlipDecision  
  & \significancemarker{\DEMNonpartisanLinearTestZeroFlipPPrecision} 
  & \DEMNonpartisanConstantTestZeroFlipDecision  
  & \significancemarker{\DEMNonpartisanConstantTestZeroFlipPPrecision} 
  & \REPNonpartisanCubicTestZeroFlipDecision  
  & \significancemarker{\REPNonpartisanCubicTestZeroFlipPPrecision} 
  & \REPNonpartisanLinearTestZeroFlipDecision  
  & \significancemarker{\REPNonpartisanLinearTestZeroFlipPPrecision} 
  & \REPNonpartisanConstantTestZeroFlipDecision  
  & \significancemarker{\REPNonpartisanConstantTestZeroFlipPPrecision} 
  \\ 

Test of homogeneous flip effect 
  & \DEMNonpartisanCubicTestHomogeneousFlipDecision  
  & \significancemarker{\DEMNonpartisanCubicTestHomogeneousFlipPPrecision} 
  & \DEMNonpartisanLinearTestHomogeneousFlipDecision  
  & \significancemarker{\DEMNonpartisanLinearTestHomogeneousFlipPPrecision} 
  &  
  &  
  & \REPNonpartisanCubicTestHomogeneousFlipDecision  
  & \significancemarker{\REPNonpartisanCubicTestHomogeneousFlipPPrecision} 
  & \REPNonpartisanLinearTestHomogeneousFlipDecision  
  & \significancemarker{\REPNonpartisanLinearTestHomogeneousFlipPPrecision} 
  & 
  & 
  \\

\midrule
$\hat{\theta}_0$ 
  & \DEMNonpartisanCubicThetaZeroEST  
  & \significancemarker{\DEMNonpartisanCubicThetaZeroPPrecision} 
  & \DEMNonpartisanLinearThetaZeroEST  
  & \significancemarker{\DEMNonpartisanLinearThetaZeroPPrecision} 
  & \DEMNonpartisanConstantThetaZeroEST  
  & \significancemarker{\DEMNonpartisanConstantThetaZeroPPrecision} 
  & \REPNonpartisanCubicThetaZeroEST  
  & \significancemarker{\REPNonpartisanCubicThetaZeroPPrecision} 
  & \REPNonpartisanLinearThetaZeroEST  
  & \significancemarker{\REPNonpartisanLinearThetaZeroPPrecision} 
  & \REPNonpartisanConstantThetaZeroEST  
  & \significancemarker{\REPNonpartisanConstantThetaZeroPPrecision} 
  \\ 
& {(\DEMNonpartisanCubicThetaZeroSE)}  
  &  
  & {(\DEMNonpartisanLinearThetaZeroSE)}  
  &  
  & {(\DEMNonpartisanConstantThetaZeroSE)}  
  &  
  & {(\REPNonpartisanCubicThetaZeroSE)}  
  &  
  & {(\REPNonpartisanLinearThetaZeroSE)}  
  &  
  & {(\REPNonpartisanConstantThetaZeroSE)}  
  &  
  \\

$\hat{\theta}_1$ 
  & \DEMNonpartisanCubicThetaOneEST  
  & \significancemarker{\DEMNonpartisanCubicThetaOnePPrecision} 
  & \DEMNonpartisanLinearThetaOneEST  
  & \significancemarker{\DEMNonpartisanLinearThetaOnePPrecision} 
  &  
  &  
  & \REPNonpartisanCubicThetaOneEST  
  & \significancemarker{\REPNonpartisanCubicThetaOnePPrecision} 
  & \REPNonpartisanLinearThetaOneEST  
  & \significancemarker{\REPNonpartisanLinearThetaOnePPrecision} 
  &  
  &  
  \\ 
& {(\DEMNonpartisanCubicThetaOneSE)}  
  &  
  & {(\DEMNonpartisanLinearThetaOneSE)}  
  &  
  &  
  &  
  & {(\REPNonpartisanCubicThetaOneSE)}  
  &  
  & {(\REPNonpartisanLinearThetaOneSE)}  
  &  
  &  
  & 
  \\

$\hat{\theta}_2$ 
  & \DEMNonpartisanCubicThetaTwoEST  
  & \significancemarker{\DEMNonpartisanCubicThetaTwoPPrecision} 
  &  
  &  
  &  
  &  
  & \REPNonpartisanCubicThetaTwoEST  
  & \significancemarker{\REPNonpartisanCubicThetaTwoPPrecision} 
  &  
  &  
  &  
  & 
  \\ 
& {(\DEMNonpartisanCubicThetaTwoSE)}  
  &  
  &  
  &  
  &  
  &  
  & {(\REPNonpartisanCubicThetaTwoSE)}  
  &  
  &  
  &  
  &  
  & 
  \\

$\hat{\theta}_3$ 
  & \DEMNonpartisanCubicThetaThreeEST  
  & \significancemarker{\DEMNonpartisanCubicThetaThreePPrecision} 
  &  
  &  
  &  
  &  
  & \REPNonpartisanCubicThetaThreeEST  
  & \significancemarker{\REPNonpartisanCubicThetaThreePPrecision} 
  &  
  &  
  &  
  &  
  \\ 
& {(\DEMNonpartisanCubicThetaThreeSE)}  
  &  
  &  
  &  
  &  
  &  
  & {(\REPNonpartisanCubicThetaThreeSE)}  
  &  
  &  
  &  
  &  
  &  
  \\

\bottomrule
\end{tabular}}
\caption*{\emph{Note.} 
The table reports estimates for the flip effect \emph{without} party labels under the cubic ($q = 3$), linear ($q = 1$), and constant ($q = 0$) specifications. 
For each party and each specification, the table reports 
estimates of partisan-voting mistakes \eqref{eq:partisan-voting-mistakes}; 
the test of zero flip effect \eqref{test:ZeroFlipEffect}; 
the test of homogeneous flip effect \eqref{test:HomogeneousFlipEffect}; 
and the estimated flip-effect coefficients.
For each party, these estimates are based on {\nNonpartisanObservationsOneParty} precinct-level observations from judicial candidates across {\nNonpartisanContests} races \emph{without} party labels.
Of these races, {\nNonpartisanContestsFlipped} exhibit a party-order flip relative to the presidential race, 
while the remaining {\nNonpartisanContestsNotFlipped} do not.
(Standard errors in parentheses. \significancemarker{0.09}$p<0.10$, 
\significancemarker{0.04}$p<0.05$, 
\significancemarker{0.009}$p<0.01$, 
\significancemarker{0.0009}$p<0.001$.)}
\end{adjustwidth}
\end{table}

We now examine two alternative specifications of the flip effect: a \emph{linear specification}, which sets $q = 1$ in \eqref{eq:f-qpoly-specification}, and a \emph{constant specification}, which sets $q = 0$.
In the absence of party designations for judicial contests,
Table \ref{tab:nonpartisan-threespecifications-summary} summarizes, by party, the results of the test of zero flip effect, the test of homogeneous flip effect, the estimated share of partisan-voting mistakes, and the flip coefficient estimates along with their standard errors and significance levels.

The linear specification yields results consistent with those from the cubic specification: for both parties, the flip effect is statistically significant, non-zero, and heterogeneous with respect to the presidential vote share.
Using the same data as in Section~\ref{se:results-no-party-labels}, but under the linear specification of the flip effect, we estimate the average proportion of partisan-voting mistakes as $\DEMNonpartisanLinearVotingMistakesESTPercentage$ for Democrats 
(p-value: $\DEMNonpartisanLinearVotingMistakesP$; 95\% confidence interval: $[\DEMNonpartisanLinearVotingMistakesCILeftPercentage, \DEMNonpartisanLinearVotingMistakesCIRightPercentage]$) 
and $\REPNonpartisanLinearVotingMistakesESTPercentage$ for Republicans 
(p-value: $\REPNonpartisanLinearVotingMistakesP$; 95\% confidence interval: $[\REPNonpartisanLinearVotingMistakesCILeftPercentage, \REPNonpartisanLinearVotingMistakesCIRightPercentage]$). 
As with the cubic specification, these findings suggest no statistically significant difference in the shares of partisan-voting mistakes between the two parties. 
Furthermore, the proportions of partisan-voting mistakes estimated under the linear and cubic specifications are not statistically different from each other.
We {\DEMNonpartisanLinearTestZeroFlipDecision} the null hypothesis \eqref{test:ZeroFlipEffect} of a zero flip effect (p-values: $\DEMNonpartisanLinearTestZeroFlipP$ for Democrats, and $\REPNonpartisanLinearTestZeroFlipP$ for Republicans).
We also {\DEMNonpartisanLinearTestHomogeneousFlipDecision} the null hypothesis \eqref{test:HomogeneousFlipEffect} of a homogeneous flip effect ($\DEMNonpartisanLinearTestHomogeneousFlipP$ for Democrats, and $\REPNonpartisanLinearTestHomogeneousFlipP$ for Republicans).

This analysis also informs our interpretation of the second- and third-degree coefficients in the cubic model. 
As shown in Tables \ref{tab:nonpartisan-cubic-detailed} and \ref{tab:nonpartisan-threespecifications-summary}, the estimates for $\theta_2$ and $\theta_3$ 
are not statistically significant for either party. 
When these higher-order terms are excluded---yielding the linear specification---the remaining coefficients become highly significant. 
This pattern implies that the flip effect is heterogeneous and monotonic in the presidential vote share and is approximately antisymmetric about $x = 0.5$.
However, such a conclusion requires analysis of both the cubic and linear specifications, as the latter enforces antisymmetry by construction. 

Finally, the cubic model also allows for a formal test of linearity of the flip effect. This involves testing the null hypothesis $H^\ell_0: \theta_2 = \theta_3 = 0$.
We {\DEMNonpartisanCubicTestLinearFlipDecision} this hypothesis for both parties, with p-values of {\DEMNonpartisanCubicTestLinearFlipP} for Democrats and {\REPNonpartisanCubicTestLinearFlipP} for Republicans. This result is consistent with  the share of partisan-voting mistakes being independent of the precinct. 

The constant specification (i.e., $q=0$ in equation \eqref{eq:f-qpoly-specification}) of the flip effect yields markedly different conclusions. In fact, by specifying the flip effect as a constant, the coefficient we are recovering is an estimate of the average flip effect (i.e., ATE).  
Using the same nonpartisan dataset employed for the cubic and linear specifications, we {\DEMNonpartisanConstantTestZeroFlipDecision} the null hypothesis of a zero flip effect (p-values: {\DEMNonpartisanConstantTestZeroFlipP} for Democrats and {\REPNonpartisanConstantTestZeroFlipP} for Republicans).
These non-significant test results, in contrast with the highly significant findings under the linear and cubic specifications, indicate that heterogeneity is a crucial feature of the flip effect. Partisan voters from both parties make voting mistakes that can be attributed to party-order flips at similar rates. 
However, because these mistakes result in reciprocal vote transfers between candidates---and given the approximately even partisan distribution of the electorate in the state---the net (unconditional) average flip effect of party-order flips is not statistically significant.

\section{Placebo test}\label{app:placebo}

\begin{table}[!t]
\linespread{1.25}\selectfont

\caption{\textbf{Placebo test with party labels (cubic specification)}}
\label{tab:partisan-cubic-detailed}

\begin{adjustwidth}{-1cm}{-1cm}
\centering
\scalebox{0.95}{%
\footnotesize
\begin{tabular}{c@{\hskip 16pt}
                r@{\hskip 8pt}
                c@{\hskip 8pt}
                r@{\hskip 8pt}
                c@{\hskip 8pt}
                r@{\hskip 0pt}
                r@{\hskip 0pt}
                r@{\hskip 0.166667em}
                r@{\hskip 0pt}
                l@{\hskip 0pt} 
                c@{\hskip 16pt}
                r@{\hskip 8pt}
                c@{\hskip 8pt}
                r@{\hskip 8pt}
                c@{\hskip 8pt}
                r@{\hskip 0pt}
                r@{\hskip 0pt}
                r@{\hskip 0.166667em}
                r@{\hskip 0pt}
                l}

\toprule
&
\multicolumn{9}{c}{\textbf{Democratic party (placebo test)}}
&
&
\multicolumn{9}{c}{\textbf{Republican party (placebo test)}} \\
\cmidrule(lr){2-10}
\cmidrule(lr){12-20}

\multicolumn{3}{l}{Partisan-voting mistakes ($\hat \tau$)}
&
\multicolumn{7}{r}{\DEMPartisanCubicVotingMistakesESTPercentage\,
{(SE: \DEMPartisanCubicVotingMistakesSEPercentage)}}
&
&
\multicolumn{9}{r}{\REPPartisanCubicVotingMistakesESTPercentage\,
{(SE: \REPPartisanCubicVotingMistakesSEPercentage)}}
\\

\multicolumn{3}{l}{Test of zero flip effect}
&
\multicolumn{7}{r}{\DEMPartisanCubicTestZeroFlipDecision,
p-value: \DEMPartisanCubicTestZeroFlipP}
&
&
\multicolumn{9}{r}{\REPPartisanCubicTestZeroFlipDecision,
p-value: \REPPartisanCubicTestZeroFlipP}
\\

\multicolumn{3}{l}{Test of homogeneous flip effect}
&
\multicolumn{7}{r}{\DEMPartisanCubicTestHomogeneousFlipDecision,
p-value: \DEMPartisanCubicTestHomogeneousFlipP}
&
&
\multicolumn{9}{r}{\REPPartisanCubicTestHomogeneousFlipDecision,
p-value: \REPPartisanCubicTestHomogeneousFlipP}
\\

\midrule
\makecell[c]{\small\textit{Coefficient}}
  & \makecell[c]{\small\textit{Estimate}}
  & \makecell[c]{\small\textit{SE}}
  & \makecell[c]{\small\textit{t}}
  & \makecell[c]{\small\textit{p-value}}
  & \multicolumn{5}{c}{\makecell[c]{\small\emph{95\% CI}}}
  &
  & \makecell[c]{\small\textit{Estimate}}
  & \makecell[c]{\small\textit{SE}}
  & \makecell[c]{\small\textit{t}}
  & \makecell[c]{\small\textit{p-value}}
  & \multicolumn{5}{c}{\makecell[c]{\small\emph{95\% CI}}} \\
\midrule

$\hat\theta_{0}$
  & \DEMPartisanCubicThetaZeroEST
  & \DEMPartisanCubicThetaZeroSE
  & \DEMPartisanCubicThetaZeroT
  & \DEMPartisanCubicThetaZeroP
  & [
  & \DEMPartisanCubicThetaZeroCILeft
  & ,
  & \DEMPartisanCubicThetaZeroCIRight
  & ]
  &
  & \REPPartisanCubicThetaZeroEST
  & \REPPartisanCubicThetaZeroSE
  & \REPPartisanCubicThetaZeroT
  & \REPPartisanCubicThetaZeroP
  & [
  & \REPPartisanCubicThetaZeroCILeft
  & ,
  & \REPPartisanCubicThetaZeroCIRight
  & ]
  \\

$\hat\theta_{1}$
  & \DEMPartisanCubicThetaOneEST
  & \DEMPartisanCubicThetaOneSE
  & \DEMPartisanCubicThetaOneT
  & \DEMPartisanCubicThetaOneP
  & [
  & \DEMPartisanCubicThetaOneCILeft
  & ,
  & \DEMPartisanCubicThetaOneCIRight
  & ]
  &
  & \REPPartisanCubicThetaOneEST
  & \REPPartisanCubicThetaOneSE
  & \REPPartisanCubicThetaOneT
  & \REPPartisanCubicThetaOneP
  & [
  & \REPPartisanCubicThetaOneCILeft
  & ,
  & \REPPartisanCubicThetaOneCIRight
  & ]
  \\

$\hat\theta_{2}$
  & \DEMPartisanCubicThetaTwoEST
  & \DEMPartisanCubicThetaTwoSE
  & \DEMPartisanCubicThetaTwoT
  & \DEMPartisanCubicThetaTwoP
  & [
  & \DEMPartisanCubicThetaTwoCILeft
  & ,
  & \DEMPartisanCubicThetaTwoCIRight
  & ]
  &
  & \REPPartisanCubicThetaTwoEST
  & \REPPartisanCubicThetaTwoSE
  & \REPPartisanCubicThetaTwoT
  & \REPPartisanCubicThetaTwoP
  & [
  & \REPPartisanCubicThetaTwoCILeft
  & ,
  & \REPPartisanCubicThetaTwoCIRight
  & ]
  \\

$\hat\theta_{3}$
  & \DEMPartisanCubicThetaThreeEST
  & \DEMPartisanCubicThetaThreeSE
  & \DEMPartisanCubicThetaThreeT
  & \DEMPartisanCubicThetaThreeP
  & [
  & \DEMPartisanCubicThetaThreeCILeft
  & ,
  & \DEMPartisanCubicThetaThreeCIRight
  & ]
  &
  & \REPPartisanCubicThetaThreeEST
  & \REPPartisanCubicThetaThreeSE
  & \REPPartisanCubicThetaThreeT
  & \REPPartisanCubicThetaThreeP
  & [
  & \REPPartisanCubicThetaThreeCILeft
  & ,
  & \REPPartisanCubicThetaThreeCIRight
  & ]
  \\

\bottomrule
\end{tabular}%
}

\caption*{\emph{Note.}
The table presents the results of a placebo test for judicial candidates in contests \emph{with} party labels under a cubic specification ($q = 3$).
The table reports estimates of partisan-voting mistakes \eqref{eq:partisan-voting-mistakes}
and results for the test of zero flip effect \eqref{test:ZeroFlipEffect}
and the test of homogeneous flip effect \eqref{test:HomogeneousFlipEffect}.
These estimates are based on {\nPartisanObservationsOneParty} precinct-level observations from Democratic judicial candidates across {\nPartisanContests} races \emph{with} party labels.
Of these races, {\nPartisanContestsFlipped} exhibit a party-order flip relative to the presidential race,
while the remaining {\nPartisanContestsNotFlipped} do not.
The Republican placebo estimates are based on observations from Republican judicial candidates in the same set of races.
}
\end{adjustwidth}
\end{table}

Table \ref{tab:partisan-cubic-detailed} presents the estimates related to the flip effect from our placebo test analysis on {\nPartisanObservationsOneParty} precinct-level observations from {\nPartisanContests} judicial races \emph{with} party labels. 
The table serves as the  placebo test counterpart to Table \ref{tab:nonpartisan-cubic-detailed}.
As expected, the estimates reported in Table~\ref{tab:partisan-cubic-detailed} provide no statistically significant evidence of a flip effect.

\newpage

\spacingset{1.25} 

\bibliography{NCElection-biblio}

@article{MillerKrosnick:POQ1998,
 author = {Miller, Joanne M. and Krosnick, Jon A.},
 journal = {The Public Opinion Quarterly},
 number = {3},
 pages = {291--330},
 publisher = {[Oxford University Press, American Association for Public Opinion Research]},
 title = {The Impact of Candidate Name Order on Election Outcomes},
 urldate = {2024-03-09},
 volume = {62},
 year = {1998}
}

@article{BonneauCann:PB2015,
 author = {Bonneau, Chris W. and Cann, Damon M.},
 journal = {Political Behavior},
 number = {1},
 pages = {43--66},
 publisher = {Springer},
 title = {Party Identification and Vote Choice in Partisan and Nonpartisan Elections},
 urldate = {2024-03-11},
 volume = {37},
 year = {2015}
}

@article{HuiHastie:JRSSB2005,
    author = {Zou, Hui and Hastie, Trevor},
    title = {Regularization and Variable Selection Via the Elastic Net},
    journal = {Journal of the Royal Statistical Society Series B: Statistical Methodology},
    volume = {67},
    number = {2},
    pages = {301-320},
    year = {2005},
    month = {04},
    issn = {1369-7412},
    doi = {10.1111/j.1467-9868.2005.00503.x},
}

@article{FriedmanHastieTibshirani:JSS2010,
 title={Regularization Paths for Generalized Linear Models via Coordinate Descent},
 volume={33},
 doi={10.18637/jss.v033.i01},
 number={1},
 journal={Journal of Statistical Software},
 author={Friedman, Jerome H. and Hastie, Trevor and Tibshirani, Rob},
 year={2010},
 pages={1–22}
}

@misc{EconML,
  author = {Battocchi, Keith and Dillon, Eleanor and Hei, Maggie and
            Lewis, Greg and Oka, Paul and Oprescu, Miruna and
            Syrgkanis, Vasilis},
  title = {{EconML}: A Python Package for ML-Based Heterogeneous Treatment Effects Estimation},
  year = {2019},
  note = {Version 0.x},
  url = {https://github.com/py-why/EconML}
}

@article{MacInnisMillerKrosnickBelowLindner:PLoS2021,
  title={Candidate name order effects in New Hampshire: Evidence from primaries and from general elections with party column ballots},
  author={MacInnis, Bo  and  Miller, Joanne M. and  Krosnick, Jon A. and  Below, Clifton and Lindner, Miriam },
  journal={PLoS ONE},
  year={2021},
  volume={16},
  url={https://api.semanticscholar.org/CorpusID:232261649}
}

@article{AugenblickNicholson:REStud2015,
    author = {Augenblick, Ned and Nicholson, Scott},
    title = "{Ballot position, choice fatigue, and voter behaviour}",
    journal = {The Review of Economic Studies},
    volume = {83},
    number = {2},
    pages = {460-480},
    year = {2015},
    month = {12},
}

@article{HoImai:JASA2006,
  title={Randomization inference with natural experiments: An analysis of ballot effects in the 2003 {C}alifornia recall election},
  author={Ho, Daniel E and Imai, Kosuke},
  journal={Journal of the American Statistical Association},
  volume={101},
  number={475},
  pages={888--900},
  year={2006},
  publisher={Taylor \& Francis}
}

@article{KingLeigh:SSQ2009,
  title={Are ballot order effects heterogeneous?},
  author={King, Amy and Leigh, Andrew},
  journal={Social Science Quarterly},
  volume={90},
  number={1},
  pages={71--87},
  year={2009},
  publisher={Wiley Online Library}
}

@article{KoppellSteen:JPol2004,
  title={The effects of ballot position on election outcomes},
  author={Koppell, Jonathan G.S. and Steen, Jennifer A.},
  journal={The Journal of Politics},
  volume={66},
  number={1},
  pages={267--281},
  year={2004},
  publisher={Cambridge University Press New York, USA}
}

@article{MeredithSalant:PoliBe2013,
  title={On the causes and consequences of ballot order effects},
  author={Meredith, Marc and Salant, Yuval},
  journal={Political Behavior},
  volume={35},
  pages={175--197},
  year={2013},
  publisher={Springer}
}

@article{McDermott:AmJPoliSci1997,
  title={Voting cues in low-information elections: Candidate gender as a social information variable in contemporary {U}nited {S}tates elections},
  author={McDermott, Monika L.},
  journal={American Journal of Political Science},
  pages={270--283},
  volume={41},
  year={1997},
  publisher={JSTOR}
}

@article{McDermott:JPol2005,
  title={Candidate occupations and voter information shortcuts},
  author={McDermott, Monika L.},
  journal={The Journal of Politics},
  volume={67},
  number={1},
  pages={201--219},
  year={2005},
  publisher={Cambridge University Press New York, USA}
}

@article{EngstromCaridas:PPAD1991,
  title={Voting for judges: Race and roll-off in judicial elections},
  author={Engstrom, Richard L. and Caridas, Victoria M.},
  journal={Political Participation and American Democracy},
  volume={92},
  pages={171--191},
  year={1991},
  publisher={Greenwood Press New York}
}

@article{VanderleeuwUtter:SSQ1993,
  title={Voter roll-off and the electoral context: A test of two theses},
  author={Vanderleeuw, James M. and Utter, Glenn H.},
  journal={Social Science Quarterly},
  pages={664--673},
  year={1993},
  volume = {74},
  publisher={JSTOR}
}

@article{WandShottsSekhon:AmPoliSciRev2001,
  title={The butterfly did it: The aberrant vote for {B}uchanan in {P}alm {B}each {C}ounty, {F}lorida},
  author={Wand, Jonathan N. and Shotts, Kenneth W. and Sekhon, Jasjeet S. and Mebane, Walter R. and Herron, Michael C. and Brady, Henry E.},
  journal={American Political Science Review},
  volume={95},
  number={4},
  pages={793--810},
  year={2001},
  publisher={Cambridge University Press}
}

@article {VanderWeele:StatMed2008,
    AUTHOR = {VanderWeele, Tyler J.},
     TITLE = {Ignorability and stability assumptions in neighborhood effects
              research},
   JOURNAL = {Statistics in Medicine},
  FJOURNAL = {Statistics in Medicine},
    VOLUME = {27},
      YEAR = {2008},
    NUMBER = {11},
     PAGES = {1934--1943},
   MRCLASS = {99-01},
  MRNUMBER = {2420353},
}

@techreport{chernozhukov2018generic,
  title={Generic machine learning inference on heterogeneous treatment effects in randomized experiments, with an application to immunization in {I}ndia},
  author={Chernozhukov, Victor and Demirer, Mert and Duflo, Esther and Fernandez-Val, Ivan},
  year={2018},
  institution={National Bureau of Economic Research}
}

@article {ChernozhukovChetverikovDemirerDufloHansenNeweyRobins:EconomJ2018,
    AUTHOR = {Chernozhukov, Victor and Chetverikov, Denis and Demirer, Mert
              and Duflo, Esther and Hansen, Christian and Newey, Whitney and
              Robins, James},
     TITLE = {Double/debiased machine learning for treatment and structural
              parameters},
   JOURNAL = {The Econometrics Journal},
  FJOURNAL = {The Econometrics Journal},
    VOLUME = {21},
      YEAR = {2018},
    NUMBER = {1},
     PAGES = {C1--C68},
   MRCLASS = {62F10 (62F12 62G05 62G20 62P20)},
  MRNUMBER = {3769544},
MRREVIEWER = {Simos\ G.\ Meintanis},
}

@article{ChiangKatoMaSasaki:JBusEconStat2022,
  title={Multiway cluster robust double/debiased machine learning},
  author={Chiang, Harold D and Kato, Kengo and Ma, Yukun and Sasaki, Yuya},
  journal={Journal of Business \& Economic Statistics},
  volume={40},
  number={3},
  pages={1046--1056},
  year={2022},
  publisher={Taylor \& Francis}
}

@article{cameron2015practitioner,
  title={A practitioner’s guide to cluster-robust inference},
  author={Cameron, A Colin and Miller, Douglas L},
  journal={Journal of human resources},
  volume={50},
  number={2},
  pages={317--372},
  year={2015},
  publisher={University of Wisconsin Press}
}

@article{HoImaiPOQ2008,
    author = {Ho, Daniel E. and Imai, Kosuke},
    title = "{Estimating causal effects of ballot order from a randomized natural experiment: The {C}alifornia alphabet lottery, 1978–2002}",
    journal = {Public Opinion Quarterly},
    volume = {72},
    number = {2},
    pages = {216-240},
    year = {2008},
    month = {05},
}

@inproceedings{OprescuSyrgkanisBattocchiHEiLewis:NeurIPS2019econml,
  title={Econ{ML}: A Machine Learning Library for Estimating Heterogeneous Treatment Effects},
  author={Oprescu, Miruna and Syrgkanis, Vasilis  and Battocchi, Kieth and Hei, Maggie and Lewis, Greg},
  booktitle={33rd Conference on Neural Information Processing Systems, Vancouver, Canada},
  pages={6},
  year={2019},
}

@article{Smith:StaSci2002,
 author = {Richard L. Smith},
 journal = {Statistical Science},
 number = {4},
 pages = {441--457},
 publisher = {Institute of Mathematical Statistics},
 title = {A Statistical Assessment of {B}uchanan's Vote in {P}alm {B}each {C}ounty},
 urldate = {2024-08-10},
 volume = {17},
 year = {2002}
}

@article{SinclairMarkMooreLavisSoldat:Nature2000electoral,
  title={An electoral butterfly effect},
  author={Sinclair, Robert C. and Mark, Melvin M. and Moore, Sean E. and Lavis, Carrie A. and Soldat, Alexander S.},
  journal={Nature},
  volume={408},
  number={6813},
  pages={665--666},
  year={2000},
  publisher={Nature Publishing Group UK London}
}

@article{Fiorina:PB2002parties,
  title={Parties and partisanship: A 40-year retrospective},
  author={Fiorina, Morris P},
  journal={Political Behavior},
  volume={24},
  pages={93--115},
  year={2002},
  publisher={Springer}
}

@book{Fiorina:Hoover2017,
  title={Unstable majorities: Polarization, party sorting, and political stalemate},
  author={Fiorina, Morris P},
  year={2017},
  publisher={Hoover Press}
}

@article{KingGelman:AmJofPoliticalScience1991,
 author = {King, Gary and Gelman, Andrew},
 journal = {American Journal of Political Science},
 number = {1},
 pages = {110--138},
 publisher = {[Midwest Political Science Association, Wiley]},
 title = {Systemic Consequences of Incumbency Advantage in {U.S.} {H}ouse Elections},
 urldate = {2024-08-11},
 volume = {35},
 year = {1991}
}

@article{CoxKatz:AmJofPoliticalSci1996,
 author = {Cox, Gary W.  and Katz, Jonathan N.},
 journal = {American Journal of Political Science},
 number = {2},
 pages = {478--497},
 publisher = {[Midwest Political Science Association, Wiley]},
 title = {Why Did the Incumbency Advantage in {U.S.} {H}ouse Elections Grow?},
 urldate = {2024-08-11},
 volume = {40},
 year = {1996}
}

@article{Dubois:LawSocietyRev1984,
 author = {Dubois, Philip L. },
 journal = {Law \& Society Review},
 number = {3},
 pages = {395--436},
 publisher = {[Wiley, Law and Society Association]},
 title = {Voting Cues in Nonpartisan Trial Court Elections: A Multivariate Assessment},
 urldate = {2024-08-11},
 volume = {18},
 year = {1984}
}

@article{Bartels:AmJofPoliticalSci2000,
 author = {Bartels, Larry M. },
 journal = {American Journal of Political Science},
 number = {1},
 pages = {35--50},
 publisher = {[Midwest Political Science Association, Wiley]},
 title = {Partisanship and Voting Behavior, 1952-1996},
 urldate = {2024-08-11},
 volume = {44},
 year = {2000}
}

@article{BraderTucker:CompPolitics2012,
 author = {Brader, Ted  and  Tucker, Joshua A.},
 journal = {Comparative Politics},
 number = {4},
 pages = {403--420},
 publisher = {Comparative Politics, Ph.D. Programs in Political Science, City University of New York},
 title = {Following the Party's Lead: Party Cues, Policy Opinion, and the Power of Partisanship in Three Multiparty Systems},
 urldate = {2024-08-11},
 volume = {44},
 year = {2012}
}

@article{CoxMorgenstern:LegislStudiesQuar1993,
 author = {Cox, Gary W.  and Morgenstern, Scott},
 journal = {Legislative Studies Quarterly},
 number = {4},
 pages = {495--514},
 publisher = {[Wiley, Comparative Legislative Research Center]},
 title = {The Increasing Advantage of Incumbency in the {U.S.} States},
 urldate = {2024-08-11},
 volume = {18},
 year = {1993}
}

@article{BelottiBuchananEzazipour:OPRE2025-Redistricting,
author = {Belotti, Pietro and Buchanan, Austin and Ezazipour, Soraya},
title = {Political Districting to Optimize the {P}olsby-{P}opper Compactness Score with Application to Voting Rights},
journal = {Operations Research},
volume = {},
number = {},
pages = {forthcoming},
year = {2025},
}

@article{CachonKaaua:MS2022-VotingResources,
author = {Cachon, G\'{e}rard P. and Kaaua, Dawson},
title = {Serving Democracy: Evidence of Voting Resource Disparity in {F}lorida},
journal = {Management Science},
volume = {68},
number = {9},
pages = {6687-6696},
year = {2022},
}

@article{CrimminsHaldermanSturt:OPRE2025-VotingMachines,
author = {Crimmins, Braden L. and Halderman, J. Alex and Sturt, Bradley},
title = {Improving the Security of {U}nited {S}tates Elections with Robust Optimization},
journal = {Operations Research},
volume = {73},
number = {1},
pages = {61-85},
year = {2025},
}

@article{FryOhlmann:Interfaces2009-VotingMachines,
author = {Fry, Michael J. and Ohlmann, Jeffrey W.},
title = {Route Design for Delivery of Voting Machines in {H}amilton {C}ounty, {O}hio},
journal = {Interfaces},
volume = {39},
number = {5},
pages = {443-459},
year = {2009},
}

@article{GordonLovettLuoReeder:MS2023=AdToneVoterTurnout,
author = {Gordon, Brett R. and Lovett, Mitchell J. and Luo, Bowen and Reeder, James C.},
title = {Disentangling the Effects of Ad Tone on Voter Turnout and Candidate Choice in Presidential Elections},
journal = {Management Science},
volume = {69},
number = {1},
pages = {220-243},
year = {2023},
}

@article{PhillipsUrbanyReynolds:JCR2008,
  title={Confirmation and the effects of valenced political advertising: A field experiment},
  author={Phillips, Joan M and Urbany, Joel E and Reynolds, Thomas J},
  journal={Journal of Consumer Research},
  volume={34},
  number={6},
  pages={794--806},
  year={2008},
  publisher={The University of Chicago Press}
}

@article{WangLewisSchweidel:MarkSci2018-PoliticalAds,
  title={A border strategy analysis of ad source and message tone in senatorial campaigns},
  author={Wang, Yanwen and Lewis, Michael and Schweidel, David A},
  journal={Marketing Science},
  volume={37},
  number={3},
  pages={333--355},
  year={2018},
  publisher={INFORMS}
}

@article{ShahmizadBuchanan:OPRE2025-Redistricting,
author = {Shahmizad, Maral and Buchanan, Austin},
title = {Political Districting to Minimize County Splits},
journal = {Operations Research},
volume = {73},
number = {2},
pages = {752-774},
year = {2025},
}

@article{SwamyKingDouglasJacobson:OPRE2023-Redistricting,
author = {Swamy, Rahul and King, Douglas M. and Jacobson, Sheldon H.},
title = {Multiobjective Optimization for Politically Fair Districting: A Scalable Multilevel Approach},
journal = {Operations Research},
volume = {71},
number = {2},
pages = {536-562},
year = {2023},
}

@article{ValidiBuchananLykhovyd:OPRE2022-Redistricting,
author = {Validi, Hamidreza and Buchanan, Austin and Lykhovyd, Eugene},
title = {Imposing Contiguity Constraints in Political Districting Models},
journal = {Operations Research},
volume = {70},
number = {2},
pages = {867-892},
year = {2022},
}

@article{GlassermanKuo:WP2025-VotingBias,
author = {Glasserman, Paul and  Kou, Steven},
title = {An Endogenous Equilibrium Poll Method to Mitigate Non-Voting Bias},
journal = {Work in progress},
volume = {},
number = {},
pages = {},
year = {2025},
doi = {}
}

@article{SchmidtBuellAlbert:MSOM2024-PollingLocations,
author = {Schmidt, Adam P. and Buell, Duncan and Albert, Laura A.},
title = {Optimal Consolidation of Polling Locations},
journal = {Manufacturing \& Service Operations Management},
volume = {26},
number = {3},
pages = {1028-1042},
year = {2024},
}

@article{SchmidtAlbert:IISE2024-DropBox,
author = {Schmidt, Adam and Albert, Laura A.},
title = {The drop box location problem},
journal = {IISE Transactions},
volume = {56},
number = {4},
pages = {424--436},
year = {2024},
publisher = {Taylor \& Francis},
}

@article{HaseltineAlbert:WP2024-VotingByMail,
      title={Voting by mail: {A} {M}arkov chain model for managing the security risks of election systems}, 
      author={Haseltine, Carmen A.  and Albert, Laura A.},
journal={ar{X}iv},
      year={2024},
pages={2410.13900},
      eprint={2410.13900},
      archivePrefix={arXiv},
      primaryClass={cs.CR},
}

@article{AutryCarterHerschlagHunterMattingly:MMS2021-Redistricting,
author = {Autry, Eric A. and Carter, Daniel and Herschlag, Gregory J. and Hunter, Zach and Mattingly, Jonathan C.},
title = {Metropolized Multiscale Forest Recombination for Redistricting},
journal = {Multiscale Modeling \& Simulation},
volume = {19},
number = {4},
pages = {1885-1914},
year = {2021},
}

@inproceedings{GurneeShmoys:ACDA2021-Redistricting,
author = {Wes Gurnee and David B. Shmoys},
title = {Fairmandering: A column generation heuristic for fairness-optimized political districting},
booktitle = {Proceedings of the 2021 SIAM Conference on Applied and Computational Discrete Algorithms (ACDA21)},
chapter = {},
year = {2021},
pages = {88-99},
}

@article{DeFordDuchinSolomon:HDSR2021-redistricting,
  title={Recombination: A family of {M}arkov chains for redistricting},
  author={DeFord, Daryl and Duchin, Moon and Solomon, Justin},
  journal={Harvard Data Science Review},
  volume={3},
  number={1},
  pages={3},
  year={2021},
  publisher={The MIT Press}
}

@article{BernsteinDuchin:NofAMS2017-gerrymandering,
  title={A formula goes to court: Partisan gerrymandering and the efficiency gap},
  author={Bernstein, Mira and Duchin, Moon},
  journal={Notices of the AMS},
  volume={64},
  number={9},
  pages={1020--1024},
  year={2017}
}

@article{HessWeaverSiegfeldtWhelanZitlau:OPRE1965-redistricting,
  title={Nonpartisan political redistricting by computer},
  author={Hess, Sidney Wayne and Weaver, JB and Siegfeldt, HJ and Whelan, JN and Zitlau, PA},
  journal={Operations Research},
  volume={13},
  number={6},
  pages={998--1006},
  year={1965},
  publisher={INFORMS}
}

@article{ShiraniMehrRothschildGoelGelman:JASA2018-BiasInPolls,
  title={Disentangling bias and variance in election polls},
  author={Shirani-Mehr, Houshmand and Rothschild, David and Goel, Sharad and Gelman, Andrew},
  journal={Journal of the American Statistical Association},
  volume={113},
  number={522},
  pages={607--614},
  year={2018},
  publisher={Taylor \& Francis}
}

@article{GelmanHullmanWlezienElliottMorris:JDM2020, 
    title={Information, incentives, and goals in election forecasts}, 
    volume={15}, 
    number={5}, 
    journal={Judgment and Decision Making}, 
    author={Gelman, Andrew and Hullman, Jessica and Wlezien, Christopher and Elliott Morris, George}, 
    year={2020}, 
    pages={863–880}
}

@article{GelmanKing:AJPS1990-Incumbency,
 author = {Andrew Gelman and Gary King},
 journal = {American Journal of Political Science},
 number = {4},
 pages = {1142--1164},
 publisher = {[Midwest Political Science Association, Wiley]},
 title = {Estimating Incumbency Advantage without Bias},
 volume = {34},
 year = {1990}
}

@article{GelmanKing:APSR1994-Redistricting,
  title={Enhancing democracy through legislative redistricting},
  author={Gelman, Andrew and King, Gary},
  journal={American Political Science Review},
  volume={88},
  number={3},
  pages={541--559},
  year={1994},
  publisher={Cambridge University Press}
}

@article{Gelman:SPP2021-PoliticalPolling,
author = {Andrew Gelman},
title = {Failure and Success in Political Polling and Election Forecasting},
journal = {Statistics and Public Policy},
volume = {8},
number = {1},
pages = {67--72},
year = {2021},
publisher = {ASA Website},
}

@article{BailArgyleBrownBumpusChenHunzakerLeeMannMerhoutVolfovsky-PNAS2018-PoliticalPolarization,
author = {Christopher A. Bail  and Lisa P. Argyle  and Taylor W. Brown  and John P. Bumpus  and Haohan Chen  and M. B. Fallin Hunzaker  and Jaemin Lee  and Marcus Mann  and Friedolin Merhout  and Alexander Volfovsky },
title = {Exposure to opposing views on social media can increase political polarization},
journal = {Proceedings of the National Academy of Sciences},
volume = {115},
number = {37},
pages = {9216-9221},
year = {2018},
doi = {10.1073/pnas.1804840115}
}

@article{Kousser-Phillips-Shor-PSRM-2018-Reform&Representation, 
author={Kousser, Thad and Phillips, Justin and Shor, Boris}, 
title={Reform and Representation: A New Method Applied to Recent Electoral Changes}, 
volume={6}, 
DOI={10.1017/psrm.2016.43}, 
number={4}, 
journal={Political Science Research and Methods}, 
year={2018}, 
pages={809–827}
}

\end{document}